\documentstyle[multicol,floats,eqsecnum,prd,aps,amsfonts,epsf]{revtex}
\newcounter{run}
\begin{document}
\draft


\title{Black hole evolution by spectral methods}
\author{Lawrence E. Kidder, Mark A. Scheel, and Saul A. Teukolsky}
\address{Center for Radiophysics and Space Research, Cornell
         University, Ithaca, New York, 14853}
\author{Eric D. Carlson and Gregory B. Cook}
\address{Department of Physics, Wake Forest University, Winston-Salem,
	 North Carolina, 27109}
\date{\today}
\maketitle

\begin{abstract}
Current methods of evolving a spacetime containing one or more black
holes are plagued by instabilities that prohibit long-term evolution.
Some of these instabilities may be due to the numerical method used,
traditionally finite differencing.  In this paper, we explore the use
of a pseudospectral collocation (PSC) method for the evolution of a
spherically symmetric black hole spacetime in one dimension using a
hyperbolic formulation of Einstein's equations.  We demonstrate that
our PSC method is able to evolve a spherically symmetric black hole
spacetime forever without enforcing constraints, even if we add
dynamics via a Klein-Gordon scalar field.  We find that, in contrast
to finite-differencing methods, black hole excision is a trivial
operation using PSC applied to a hyperbolic formulation of Einstein's
equations.  We discuss the extension of this method to three spatial
dimensions.
\end{abstract}



\begin{multicols}{2}
\section{Introduction and summary}

A major thrust of research in classical general relativity in the past
decade has been to devise algorithms to solve Einstein's equations
numerically.  Despite advances in our analytic understanding of
general relativity, we still do not know what all the features of the
theory really are. Numerical solutions will continue to provide fresh
insights into the theory as they have in the past, for example,
critical behavior in black hole formation \cite{Choptuik1993} and the
formation of toroidal black holes \cite{Shapiro1995}.

New urgency has been injected into numerical relativity by the
imminent deployment of LIGO. The prime target for LIGO is coalescence
of binary neutron star and black hole systems. The waveform is
reasonably well predicted by the post-Newtonian approximation when the
binary components are at large separation. However, extracting the
most important physics requires us to be able to deal with fully
non-linear general relativity as the system spirals together and
coalesces. Moreover, a number of people believe that there is a
significant event rate for the coalescence of massive black hole
systems ($\sim 20M_\odot$) \cite{Zwart2000}.  In this case, LIGO is
most sensitive to waves emitted from the strong field regime.  Indeed,
without some theoretical guidance as to what to expect from this
regime, it is possible we may miss these events entirely
\cite{Flanagan1998}.

However, the goal of developing a general algorithm that can solve
Einstein's equations for two black holes has remained elusive.  All
attempts to date have been plagued by instabilities.  These
instabilities are caused by an interplay of three factors: (1)
Einstein's equations are an overdetermined system, with the evolution
equations subject to constraints. So if, for example, you choose to
solve only the evolution equations, then there can be unstable
solutions that are in fact solutions of the evolution equations, but
do not satisfy the constraints. Small numerical errors may cause these
solutions to appear and swamp the true solution
(``constraint-violating modes'').  (2) The coordinate freedom inherent
in the theory means that it is very easy to impose coordinate
conditions that lead to numerical instabilities (``gauge modes''). (3)
Experience has shown that the kind of boundary conditions we choose
and how we implement them can affect the stability of an algorithm
enormously.

Similar instabilities have hampered efforts to solve the related
problem of binary neutron stars; only very recently
\cite{Landry2000,Shibata2000} has there been some success in finding
stable algorithms.  However, black hole evolutions face an additional
obstacle that is absent in the case of neutron stars: for neutron
stars the gravitational field is everywhere regular, but for black
holes one must somehow deal with the physical singularity that lurks
inside each hole.

There are two main approaches for handling these singularities. The
first is to use gauge conditions (e.g., maximal slicing) that avoid
the singularities altogether. Such conditions, however, lead to large
gradients in the gravitational field variables near the horizon.
These grow exponentially in time and ultimately cannot be resolved by
the numerical evolution, causing the code to crash.  The alternative
approach is to excise the region containing the singularity from the
computational domain and evolve only the exterior region.  If the
excision boundary is placed inside the horizon of the black hole,
causality assures us that we do not need to impose a physical boundary
condition there.

However, black hole excision is only known to be mathematically
well-posed if the evolution equations are hyperbolic with
characteristic speeds less than or equal to $c$. In this case, the
structure of the equations guarantees that even unphysical modes
present in the solution (gauge modes, constraint-violating modes)
behave causally and cannot propagate out of the horizon. For many
representations of general relativity such as the usual ADM \cite{ADM}
formulation, the evolution equations are of no mathematical type for
which well-posedness has been proven, so the suitability of these
formulations for black hole excision must be determined empirically on
a case-by-case basis. It is in part for this reason that much
attention has been recently focused on hyperbolic representations of
Einstein's equations
\cite{frittelli_reula94,choquet_york95,abrahams_etal95,%
bona_masso95b,mvp96,frittelli_reula96,friedrich96,estabrook_etal97,%
anderson_etal98,Alcubierre1999,Frittelli1999,anderson_york99}.

It is still unclear whether hyperbolic formulations are
computationally advantageous. However, it has been shown that the
formulation of the evolution equations can affect stability. For
example, some instabilities can be eliminated by changing from one
formulation of Einstein's equations to another \cite{baumgarte99} or
by modifying the evolution equations to change the spectrum of
unphysical modes \cite{scheel_etal97b}.

Nevertheless, it is likely that many instabilities encountered in
practice are due to the numerical implementation of the evolution
equations and of the boundary conditions.  Even for well-understood
systems, it is far from guaranteed that any given numerical
approximation to the continuum equations will be stable. The
well-known Courant instability is a trivial example.  Hence it is
prudent to explore alternative numerical methods that may offer a
shorter path to the goal of long-term stability.

Traditionally, black hole spacetimes have been evolved using
finite-difference (FD) methods. Current FD codes for evolving black
hole spacetimes with excised horizons are mostly based on a numerical
technique known as causal differencing
\cite{seidel_suen92,alcubierre94,scheel_etal97a,bbhprl98a,Gundlach1999,%
Lehner2000},
which allows one to update the fundamental variables in time while
avoiding numerical problems associated with superluminal grid
speeds. This technique has been used successfully to propagate an
excised hole across a grid, even when grid points fall into or emerge
from the horizon \cite{bbhprl98a}.

However, causal differencing is complicated because it requires
interpolation, and it has to deal with points ``missing'' in some
irregular fashion near the excision boundary.  The FD operator that
performs the interpolation depends on the shape of the excision
boundary. Furthermore, even for a given excision boundary and a given
target point, the operator is not unique. Hence one must construct a
large number of interpolation operators, each of which involves some
arbitrary choice, in order to perform interpolations on the entire
grid.  It is only by trial and error that one finds operators that
result in a stable evolution scheme.  For simulations of a single
spherical black hole on a Cartesian three-dimensional grid, we have
found a case in which changing a single interpolation operator that is
used only for a single target point on the grid makes the difference
between a stable and an unstable code.

Another limitation of FD methods is the difficulty of imposing
boundary conditions at the {\it outer} boundary of the calculation.
There are two aspects to this problem: First, one must formulate a
procedure for handling the boundary, such as imposing an analytic
condition (e.g., a Sommerfeld condition) on the fundamental variables,
matching to a wave perturbation described by the Zerilli equation
\cite{Abrahams1998}, or matching to a characteristic evolution code
that propagates the solution out to null infinity
\cite{Bishop1996,bbhprl98b,Bishop1998}. Second, one must construct a
FD approximation of either the analytic boundary condition or the
matching condition. It can be difficult to find such an approximation
that yields a stable evolution.

In this paper we explore an alternative computational strategy: a
pseudospectral collocation (PSC) scheme.  In our PSC evolution scheme,
the solutions to a set of hyperbolic differential equations are
approximated as series expansions in a set of orthogonal basis
functions (e.g., Chebyshev polynomials) in space, and these
coefficients are integrated forward in time using the method of lines.
PSC has three important advantages over FD: (1) No ad hoc
interpolation operators are required to determine field values at an
arbitrary point because the solution provided by PSC is an analytic
function given everywhere on the computational domain. (2) Boundary
conditions are imposed directly on the basis functions, with no
approximations, in a straightforward manner.  (3) For smooth
solutions, PSC will converge to the actual solution exponentially as
the number of basis functions is increased. A FD solution, on the
other hand, never converges faster than algebraically with the number
of grid points.  Thus, for a given accuracy, PSC requires far less CPU
time and memory than FD methods.

PSC has been applied successfully to solve problems in many fields,
including fluid dynamics, meteorology, seismology, and relativistic
astrophysics (cf.\
\cite{Boyd1989,Canuto1988,Fornberg1996,Bonazzola1999}).  For example,
PSC has been applied successfully to model stellar core collapse
\cite{Novak2000} and construct equilibrium sequences of irrotational
binary neutron stars \cite{Bonazzola1999a}.  For black hole spacetimes
PSC has been applied successfully to solve initial data for the
standard field equations \cite{Kidder2000} and the conformal field
equations \cite{Frauendiener1999}, to find apparent horizons
\cite{Gundlach1998}, to solve the shift vector equation for a Kerr
black hole \cite{Bonazzola1996}, and to evolve Einstein's equations in
null quasi-spherical coordinates \cite{Bartnik2000}.

Here we evolve a spherically symmetric black hole spacetime in one
spatial dimension by applying PSC methods to a hyperbolic formulation
of Einstein's equations, the ``Einstein-Christoffel'' (EC) system
\cite{anderson_york99}. A hyperbolic formulation provides a
well-defined prescription for imposing boundary conditions.  At a
boundary, the fundamental fields can be decomposed into characteristic
fields, which in the case of the EC system propagate either along the
light cone or normal to the spatial foliation.  Boundary conditions
are imposed on the incoming (with respect to the computational domain)
characteristic fields, but not on the outgoing fields. After imposing
the boundary condition, the fundamental fields are reconstituted from
the characteristic decomposition.  If the excision boundary is placed
inside the event horizon of a black hole, then (for appropriate
coordinate systems) all characteristic fields are outgoing at this
boundary, so no boundary conditions are needed there.  Thus, black
hole excision is a trivial operation.  At the outer boundary of the
domain, boundary conditions corresponding to no incoming radiation can
be imposed on the incoming characteristic fields.

We find that our PSC method is able to evolve a spherically symmetric
black hole spacetime forever. Furthermore, we find that the solution
converges exponentially to the exact solution as the number of basis
functions is increased.  We discuss the time-stepping algorithms,
outer boundary conditions, and gauge conditions required for stable
evolution and how these depend on the particular slicing of the
Schwarzschild geometry we wish to reproduce.  We also show that our
PSC method can handle dynamics by evolving a black hole spacetime
containing a scalar field.  Finally, we outline a strategy for
applying this method to two black holes in three spatial dimensions
using multiple domains, and we present tests of this domain
decomposition idea in spherical symmetry.

In Sec.\ \ref{sec:Evolution-equations} we list the basic equations we
use to evolve a spherically symmetric spacetime, boundary conditions,
gauge conditions, initial data, and diagnostics. In Sec.\
\ref{sec:Pseud-coll-meth} we introduce PSC and describe our numerical
methods. In Sec.\ \ref{sec:Numerical-results} we present numerical
evolutions of single black hole spacetimes in spherical symmetry.
Finally, in Sec.\ \ref{sec:Discussion} we discuss our results and the
generalization of our methods to multiple dimensions and to binary
black hole spacetimes.


\section{Basic equations}
\label{sec:Evolution-equations}

\subsection{Einstein-Christoffel system}

We adopt the ``Einstein-Christoffel'' (EC) hyperbolic representation
of Einstein's equations \cite{anderson_york99}. In this formulation
the evolution equations are written in first-order 
\\ \mbox{} \hrulefill \\
symmetric hyperbolic form, and all characteristic curves are directed
either along the light cone or normal to the spatial foliation. The EC
system takes as fundamental quantities the familiar three-metric and
extrinsic curvature plus only 18 additional ``connection'' variables
(in three spatial dimensions). Unlike some hyperbolic representations
of general relativity, the EC system requires no derivatives of the
stress-energy tensor.

We write the metric in the usual 3+1 form
\begin{equation}
ds^2 = -N^2 dt^2 + g_{ij}(dx^i + \beta^i dt)(dx^j + \beta^j dt),
\end{equation}
where $g_{ij}$ is the three-metric, $\beta^i$ is the shift vector, and
$N$ is the lapse function.  In the EC formulation it is not the lapse
function $N$ that is freely specifiable; instead, one arbitrarily
prescribes the densitized lapse function $\alpha$, defined by
\begin{equation}
\alpha \equiv \frac{N}{\sqrt{g}} ,
\end{equation}
where $g$ is the determinant of the three-metric. The use of a
densitized lapse does not fix the temporal gauge freedom in any way:
one can in principle obtain any lapse function $N$ by an appropriate
choice of $\alpha$.

To write down the EC evolution equations, first define the new
variables
\begin{equation}
f_{k i j} \equiv \Gamma_{(i j) k} + g_{k i}g^{l m}\Gamma_{[l j] m} +
g_{k j}g^{l m}\Gamma_{[l i] m},
\end{equation}
where $\Gamma^k_{~ij}$ is the affine connection associated with
$g_{ij}$, and parentheses and brackets denote symmetrization and
antisymmetrization, respectively. The quantities $f_{k i j}$ will be
taken as fundamental variables along with $g_{ij}$ and the extrinsic
curvature $K_{ij}$.

The EC evolution equations can be written in the form
\end{multicols}
\begin{mathletters} 
\label{eq:3DEC}
\begin{eqnarray}
\widehat{\partial_0}{g_{i j}} = && -2 N K_{i j}, \\
\widehat{\partial_0}{K_{i j}} + N g^{k l} \partial_l f_{k i j} = && N \{ 
g^{k l}(K_{k l} K_{i j} - 2 K_{k i} K_{l j})
+ g^{k l} g^{m n} (4 f_{k m i} f_{[ln{]}j} + 4 f_{k m[n} f_{l{]}i j}
- f_{i k m} f_{j l n} \nonumber \\ && \mbox{}
+ 8 f_{(i j)k}f_{[ln{]}m} + 4 f_{k m (i} f_{j) l n} - 8 f_{k l i} f_{m n j}
+ 20 f_{k l (i} f_{j) m n} - 13 f_{i k l} f_{j m n}) \nonumber \\ && \mbox{}
- \partial_i \partial_j \ln \alpha 
- (\partial_i \ln \alpha)(\partial_j \ln \alpha)
+ 2 g_{i j} g^{k l} g^{m n} (f_{k m n} \partial_l \ln \alpha
- f_{k m l} \partial_n \ln \alpha) \nonumber \\ && \mbox{}
+ g^{k l} [(2 f_{(i j) k} - f_{k i j})\partial_l \ln \alpha
+ 4f_{k l(i}\partial_{j)} \ln \alpha - 3(f_{i k l}\partial_j \ln \alpha 
+ f_{j k l}\partial_i \ln \alpha) {]} \nonumber \\ && \mbox{}
- 8\pi S_{ij} + 4\pi g_{i j} T \},\\
\widehat{\partial_0}{f_{k i j}} + N \partial_k K_{i j} = && N \{ 
g^{m n} [4 K_{k (i} f_{j) m n} - 4 f_{m n (i} K_{j)k}
+ K_{i j}(2 f_{m n k} - 3 f_{k m n}){]} \nonumber \\ && \mbox{}
+ 2 g^{m n} g^{p q}[K_{m p} (g_{k (i} f_{j)q n} - 2 f_{q n(i} g_{j)k})   
+ g_{k (i} K_{j)m} (8 f_{n p q} - 6 f_{p q n}) \nonumber \\ &&\mbox{}
+ K_{m n} (4 f_{p q(i} g_{j)k} - 5 g_{k (i} f_{j)p q}){]} 
- K_{i j} \partial_ k \ln \alpha \nonumber \\ && \mbox{}
+ 2 g^{m n} (K_{m (i} g_{j) k} \partial_n \ln \alpha
- K_{m n} g_{k (i} \partial_{j)} \ln \alpha)
+ 16 \pi g_{k (i} J_{j)} \}.
\end{eqnarray}
\end{mathletters}
Here the symbol $\widehat{\partial_0}$ is the time derivative operator
normal to the spatial foliation, defined by
\begin{equation}
\widehat{\partial_0} \equiv \partial_t - \pounds_\beta,
\end{equation}
where $\pounds$ denotes a Lie derivative.  The matter terms are
\begin{multicols}{2} 
\begin{mathletters}
\label{eq:MatterTermDefinitions}
\begin{eqnarray}
\rho   \equiv && n^\mu n^\nu T_{\mu\nu},\\
T      \equiv && {}^{(4)}g^{\mu\nu} T_{\mu\nu},\\
J_{i}  \equiv && -n^\mu \gamma^{~\nu}_i T_{\mu\nu},\\
S_{ij} \equiv && \gamma^{~\nu}_i \gamma^{~\mu}_j T_{\mu\nu},
\end{eqnarray}
\end{mathletters}
where $T_{\mu\nu}$ is the stress-energy tensor, ${}^{(4)}g_{\mu\nu}$
is the four-metric, $n^\mu$ is the unit normal to the spatial
foliation, and $\gamma^{~\nu}_i$ is the spatial projection operator
$n^\nu n_i+{}^{(4)}g^{~\nu}_i$.

Note that for each $\{i,j\}$ pair, the evolution
equations~(\ref{eq:3DEC}) can be written in the form (dropping the $i$
and $j$ indices on $g_{ij}$, $K_{ij}$, and $f_{kij}$)
\begin{mathletters}
\begin{eqnarray}
\label{Simple3DEC_ijpair}
\widehat{\partial_0} g \equiv && -2 N K,\\ \hline \nonumber \\
\widehat{\partial_0} K + N g^{k l} \partial_l f_k \equiv && R,\\
\widehat{\partial_0} f_k + N \partial_k K \equiv && S,
\end{eqnarray}
\end{mathletters}
where $R$ and $S$ are nonlinear terms that contain no derivatives of
the fundamental variables (but may contain spatial derivatives of the
arbitrary gauge function $\alpha$).  Except for the right-hand sides,
equations~(\ref{Simple3DEC_ijpair}) are just $\Box g = 0$ written in
first order form.  Thus one can think of the EC system~(\ref{eq:3DEC})
as a set of six (one for each $\{i,j\}$ pair) coupled quasilinear
scalar wave equations with nonlinear source terms.

A solution of the evolution equations~(\ref{eq:3DEC}) is not a solution
to Einstein's equations unless twenty-two constraints are also
satisfied.  These are the Hamiltonian constraint
\end{multicols}
\begin{mathletters}
\label{eq:3DEC_constraints}
\begin{eqnarray}
{\cal C} \equiv &&
g^{i j} g^{k l} \{ 2 (\partial_k f_{i j l} - \partial_i f_{j k l}) 
+ K_{i k} K_{j l} - K_{i j} K_{k l} 
+ g^{m n} [f_{i k m}  (5 f_{j l n} - 6 f_{l j n}) 
+ 13 f_{i k l} f_{j m n} \nonumber \\ && \mbox{}
+ f_{i j k} (8 f_{m l n} - 20 f_{l m n}) ]\} + 16\pi\rho = 0,
\end{eqnarray}
the three momentum constraints
\begin{eqnarray}
{\cal C}_i \equiv && g^{k l} \{ g^{m n} [ K_{i k} (3 f_{l m n} - 2 f_{m n l}) 
- K_{k m} f_{i l n}] 
+ \partial_i K_{k l} - \partial_k K_{i l}\} + 8 \pi J_i = 0,
\end{eqnarray}
and the eighteen constraints
\begin{equation}
{\cal C}_{kij} \equiv \partial_k g_{i j} - 2 f_{k i j}
+ 4 g^{l m} (f_{l m(i}  g_{j) k} - g_{k (i} f_{j)l m}) = 0
\end{equation}
\end{mathletters}
\begin{multicols}{2} \noindent
that relate $f_{kij}$ to spatial derivatives of the three-metric.  If
the constraints are satisfied for the initial data, they are preserved
by the evolution equations for all time.  As for any formulation of
Einstein's equations, however, numerical approximations may spoil this
constraint-preserving property.


\subsection{Reduction to spherical symmetry}

The most general spherically symmetric metric can be written in the
form
\begin{eqnarray}
ds^2 = && -N^2 dt^2 + g_{r r}(dr + \beta^r dt)^2 \nonumber \\ &&
+ g_T r^2 (d\theta^2 + \sin^2 \theta d \phi^2),
\end{eqnarray}
where the transverse metric component is defined by
\begin{equation}
g_T \equiv \frac{g_{\theta \theta}}{r^2} = 
\frac{g_{\phi \phi}}{r^2 \sin^2 \theta}.
\end{equation}
The two nonvanishing independent components of the extrinsic curvature
are $K_{rr}$ and the transverse extrinsic curvature
\begin{equation}
K_T \equiv \frac{K_{\theta \theta}}{r^2} =
\frac{K_{\phi \phi}}{r^2 \sin^2 \theta}.
\end{equation}
The nonvanishing components of $f_{kij}$ are $f_{rrr}$, the transverse
component
\begin{equation}
f_{rT} \equiv \frac{f_{r \theta \theta}}{r^2} =
\frac{f_{\theta \theta r}}{2 r^2} = \frac{f_{r \phi \phi}}{r^2 \sin^2 \theta}
= \frac{f_{\phi \phi r}}{2 r^2 \sin^2 \theta},
\end{equation}
and the additional components
\begin{mathletters}
\label{f_extra_components}
\begin{eqnarray}
f_{r r \theta} = && g_{r r} \cot \theta, \\
f_{\theta \theta \theta} = && 2 r^2 g_T \cot \theta, \\
f_{\theta \phi \phi} = && r^2 g_T \sin \theta \cos \theta, \\
f_{\phi \theta \phi} = && r^2 g_T \sin \theta \cos \theta.
\end{eqnarray}
\end{mathletters}
The evolution equations for these additional
components~(\ref{f_extra_components}) are automatically obeyed if the
evolution equations for the metric are satisfied.  We therefore do not
treat these quantities as independent variables, and wherever they
appear in the equations we replace them with the appropriate metric
components using equations~(\ref{f_extra_components}).  Of course, all
angular dependence due to these terms drops out.

We therefore take as fundamental variables the six quantities
$g_{rr}$, $g_T$, $K_{rr}$, $K_T$, $f_{rrr}$, and
$f_{rT}$. Using~(\ref{eq:3DEC}), we obtain the following evolution
equations for these variables:
\end{multicols} \pagebreak
\begin{mathletters}
\label{eq:1DEC}
\begin{eqnarray}
\partial_t g_{rr} - \beta^r \partial_r g_{rr}  = && - 2 N K_{rr} 
+ 2 g_{rr} \partial_r \beta^r,\\
\partial_t g_T - \beta^r \partial_r g_T = && - 2 N K_T 
+ 2 \frac{\beta^r}{r} g_T,\\
\partial_t K_{rr} - \beta^r \partial_r K_{rr}  
+ \frac{N}{g_{rr}} \partial_r f_{rrr} 
= && N \left[ 2 f^r_{~rr} \left(f^r_{~rr} + \frac{1}{r} 
- \frac{4 f_{rT}}{g_T} \right) - \frac{6}{r^2} 
+ K_{rr} \left( 2 \frac{K_T}{g_T} 
- K^r_{~r} \right) - 6 \left(\frac{f_{rT}}{g_T}\right)^2 
- \partial^2_r \ln \tilde{\alpha} \right. \nonumber \\ && \left. \mbox{}
- (\partial_r \ln \tilde{\alpha})^2
+ \left(\frac{4}{r} - f^r_{~rr} \right)\partial_r \ln \tilde{\alpha} \right]
+ 2 K_{rr} \partial_r \beta^r + 4 \pi N (T g_{rr} - 2 S_{rr}),\\
\partial_t K_T - \beta^r \partial_r K_T + \frac{N}{g_{rr}} \partial_r f_{rT}
= && N \left( K_T K^r_{~r} + \frac{1}{r^2} 
- \frac{2f_{rT}^2}{g_{rr}g_T} 
- \frac{f_{rT}}{g_{rr}} \partial_r \ln \tilde{\alpha} \right)
+ \frac{2 \beta^r}{r} K_T,\\\nonumber \\
\partial_t f_{rrr} - \beta^r \partial_r f_{rrr} + N \partial_r K_{rr}
= && N \left[ 4 g_{rr} \frac{K_T}{g_T} \left(3 \frac{f_{rT}}{g_T}
- f^r_{~rr}  + \frac{2}{r} - \partial_r \ln \tilde{\alpha} \right) 
- K_{rr} \left(10 \frac{f_{rT}}{g_T} + f^r_{~rr} - \frac{2}{r} 
+ \partial_r \ln \tilde{\alpha} \right) \right] \nonumber \\ && \mbox{} 
+ 3 f_{rrr} \partial_r \beta^r 
+ g_{rr} \partial^2_r \beta^r + 16 \pi N J_r g_{rr},\\ \nonumber \\
\partial_t f_{rT} - \beta^r \partial_r f_{rT} + N \partial_r K_T 
= && N \left[ K_T \left( 2 \frac{f_{rT}}{g_T} - f^r_{~rr} 
- \partial_r \ln \tilde{\alpha} \right) \right] 
+ \left( \partial_r \beta^r
+ \frac{2 \beta^r}{r}  \right) f_{rT}.
\end{eqnarray}
\end{mathletters}
Here
\begin{equation}
\tilde {\alpha} \equiv \alpha r^2 \sin \theta
= \frac{N}{g_T \sqrt{g_{rr}}},
\end{equation}
and we have explicitly included the terms involving the Lie derivative
of the shift vector.

The six fundamental variables obey four constraints that can be
obtained from~(\ref{eq:3DEC_constraints}):
\begin{mathletters}
\label{eq:1DEC:constraints}
\begin{eqnarray}
{\cal C} \equiv && \frac{\partial_r f_{rT}}{g_{rr} g_T} - \frac{1}{2 r^2 g_T}
+ \frac{f_{rT}}{g_{rr} g_T} \left( \frac{2}{r} + \frac{7 f_{rT}}{2 g_T}
- f^r_{~rr}\right) 
- \frac{K_T}{g_T} \left( K^r_{~r} + \frac{K_T}{2 g_T} \right)
+ 4 \pi \rho = 0, \label{eq:hamcon} \\
{\cal C}_r \equiv && \frac{\partial_r K_T}{g_T} + \frac{2 K_T}{r g_T}
- \frac{f_{rT}}{g_T} \left(K^r_{~r} + \frac{K_T}{g_T} \right)
+4 \pi J_r = 0, \label{eq:momcon} \\
{\cal C}_{rrr} \equiv && \partial_r g_{rr} 
+ \frac{8 g_{rr} f_{rT}}{g_T} - 2 f_{rrr}  = 0,
\label{eq:grrcon} \\
{\cal C}_{rT} \equiv && \partial_r g_T + \frac{2 g_T}{r} - 2 f_{rT} 
= 0.\label{eq:gttcon}
\end{eqnarray}
\end{mathletters}
\begin{multicols}{2}\noindent
We do not explicitly solve the constraints during our evolution, but
instead we use them as error estimators.


\subsection{Boundary conditions}\label{sec:Boundary-conditions}

Boundary conditions are imposed on the above evolution
equations~(\ref{eq:1DEC}) via characteristic decomposition.  Consider
a first-order symmetrizable hyperbolic system
\begin{equation}
\partial_t U + A^i \partial_i U = R,
\end{equation}
where $U$ is the vector of variables, $R$ is a vector, and the three
$A^i$ are matrices. Then for a particular unit vector $\xi_i$, the
solutions $U_c$ to the eigenvalue problem
\begin{equation}
A^i \xi_i U_c = v_c U_c
\end{equation}
define the characteristic fields normal to the direction $\xi_i$, and
the eigenvalues $v_c$ define the characteristic speeds of these
fields. Each of the characteristic fields $U_c$ can be thought of as a
plane wave solution moving in the direction $\xi_i$ with speed $v_c$.
One is allowed to impose boundary conditions only on characteristic
fields that propagate into the computational domain, but not on fields
that propagate out of the domain.

For the evolution equations~(\ref{eq:1DEC}), the characteristic
fields in the radial direction ($\xi_r = \sqrt{g_{rr}}$) are
\begin{mathletters}
\begin{eqnarray}
U^0_r \equiv g_{rr} &\qquad& (v_c = -\beta^r),\\ 
U^0_t \equiv g_T    &\qquad& (v_c = -\beta^r),\\ 
U^\pm_r \equiv K_{rr} \pm \frac{f_{rrr}}{\sqrt{g_{rr}}} &\qquad& 
(v_c  = -\beta^r \pm \tilde{\alpha} g_T),\\ 
U^\pm_T \equiv K_T \pm \frac{f_{rT}}{\sqrt{g_{rr}}} &\qquad&
(v_c  = -\beta^r \pm \tilde{\alpha} g_T).
\end{eqnarray}
\end{mathletters}
The characteristic speeds of the metric variables correspond to
propagation along the timelike normal to the foliation and the
characteristic speeds of the other quantities correspond to
propagation along the light cone.  Thus, if the inner boundary of our
domain moves along a spacelike trajectory, all characteristic speeds
are negative (with respect to $r$) there, so no boundary conditions
need to be imposed.  At the outer boundary, boundary conditions are
imposed only on those quantities with negative characteristic speeds
(usually $U^0_r$, $U^0_t$, $U^-_r$, and $U^-_T$).

We have experimented with three types of outer boundary conditions. The
first, which we call the freezing boundary condition, is
\begin{equation}
\label{eq:ConstantOuterBoundary}
\partial_t U_c = 0
\end{equation}
applied to all incoming characteristic fields $U_c$. This corresponds to
no incoming radiation at the boundary; however, for nonlinear or
inhomogeneous problems this is not strictly correct unless the
boundary is at infinity.

The second is the Robin condition
\begin{equation}
\label{eq:Robin}
\partial_r [r^n (U-U_\infty)] = 0,
\end{equation}
which assumes that $U$ behaves like
\begin{equation}
U_\infty + \frac{\text{const}}{r^n}
\end{equation}
at large $r$. For a given incoming variable $U_c$, appropriate values
of the parameters $U_\infty$ and $n$ can be found from the analytic
representations of the Schwarzschild geometry in
section~\ref{sec:Time-indep-slic} below.

Finally, the constraints~(\ref{eq:1DEC:constraints}) can be used to
derive mixed Neumann-Dirichlet boundary conditions for four of the
characteristic fields: Equations~(\ref{eq:grrcon})
and~(\ref{eq:gttcon}) can be used directly as boundary conditions on
$U^0_r$ and $U^0_t$, and Equations~(\ref{eq:hamcon})
and~(\ref{eq:momcon}) can be combined to yield boundary conditions on
$U^-_T$ and $U^+_T$. We use only three of these boundary
conditions---the ones for $U^0_r$, $U^0_t$, and $U^-_T$---because
$U^+_T$ is outgoing at the outer boundary and therefore needs no
boundary condition there.  Similarly, in three spatial dimensions the
constraints can be used to derive twenty-two relations among the
thirty characteristic fields, some of which can be used as boundary
conditions on the incoming fields.

We describe our numerical implementation of boundary conditions in
section~\ref{sec:Appl-bound-cond}.


\subsection{Coordinate systems}
\label{sec:Time-indep-slic}

It is convenient to choose a coordinate system in which the
Schwarzschild geometry is time-independent.  Furthermore, since we
wish to include the apparent horizon in our computational domain, we
must choose coordinates such that the spacelike slices labeled by
constant values of coordinate $t$ penetrate the horizon and are
nonsingular there.  Here we list several coordinate systems that
satisfy these properties.

\subsubsection{Kerr-Schild coordinates}
In this coordinate system, also referred to as ingoing
Eddington-Finkelstein coordinates \cite{marsa96}, ingoing null rays
have unit coordinate speed. In addition, the radial coordinate $r$ is
chosen such that $4\pi r^2$ is the surface area of a sphere at that
radius.  In this coordinate system the Schwarzschild solution takes
the form
\begin{mathletters}
\label{kerr-schild}
\begin{eqnarray}
g_{rr} = && 1 + \frac{2M}{r},\\
g_T = && 1,\\
\tilde{\alpha} = && \left(1 + \frac{2M}{r} \right)^{-1},
\label{kerr-schild-lapse}\\
\beta^r = && \frac{2M}{r} \left(1 + \frac{2M}{r} \right)^{-1},
\label{kerr-schild-shift}\\
K_{rr} = && - \frac{2M}{r^2} \left(1 + \frac{M}{r}\right)
\left(1 + \frac{2M}{r}\right)^{-1/2},\\
K_T = && \frac{2M}{r^2} \left(1 + \frac{2M}{r}\right)^{-1/2},\\
f_{rrr} = && \frac{1}{r} \left(4 + \frac{7M}{r}\right),\\
f_{rT} = && \frac{1}{r},
\end{eqnarray}
\end{mathletters}
where $M$ is the mass of the hole. The event horizon is coincident
with the apparent horizon and is located at $r=2M$.

\subsubsection{Painlev\'{e}-Gullstrand coordinates}
In this coordinate system
\cite{Gundlach1999,painleve21,gullstrand22,Martel2000} the spatial
three-metric is flat and the Schwarzschild solution is particularly
simple:
\begin{mathletters}
\label{pain-gull}
\begin{eqnarray}
g_{rr} = && 1,\\
g_T = && 1,\\
\tilde{\alpha} = && 1,\\
\beta^r = && \sqrt{\frac{2M}{r}},\\
K_{rr} = && - \sqrt{\frac{M}{2r^3}},\\
K_T = && \sqrt{\frac{2M}{r^3}},\\
f_{rrr} = && \frac{4}{r},\\
f_{rT} = && \frac{1}{r}.
\end{eqnarray}
\end{mathletters}
The horizon is again located at $r=2M$.

\subsubsection{Harmonic time slicing, areal radial coordinates}
If one requires the time coordinate to satisfy $\Box t = 0$, the
radial coordinate $r$ to correspond to the areal radius, and the
coordinate system to be regular at the horizon, then the Schwarzschild
solution takes the form \cite{bona_masso88,cook_scheel97}
\begin{mathletters}
\label{harmonic_time}
\begin{eqnarray}
g_{rr} = && \left(1 + \frac{2M}{r}\right) \left(1 + \frac{4M^2}{r^2}\right),\\
g_T = && 1,\\
\tilde{\alpha} = && \left(1 + \frac{2M}{r}\right)^{-1}
\left(1 + \frac{4M^2}{r^2}\right)^{-1},\\
\beta^r = && \frac{4\tilde{\alpha}M^2}{r^2} ,\\
K_{rr} = && - \frac{4M^2}{r^3} \sqrt{\tilde{\alpha}} \left(2
+ \frac{3M}{r} +\frac{4M^2}{r^2} + \frac{4M^3}{r^3}\right),\\
K_T = && \frac{4M^2}{r^3} \sqrt{\tilde{\alpha}},\\
f_{rrr} = && \frac{4}{r} + \frac{7M}{r^2} + \frac{12M^2}{r^3}
+ \frac{20M^3}{r^4},\\
f_{rT} = && \frac{1}{r}.
\end{eqnarray}
\end{mathletters}
The horizon is at $r=2M$.

\subsubsection{Fully harmonic coordinates}
The Schwarzschild solution can also be written in a coordinate system
where all coordinates satisfy $\Box x^\mu = 0$ and are regular
at the event horizon \cite{bona_masso88,cook_scheel97}:
\begin{mathletters}
\label{full_harmonic}
\begin{eqnarray}
g_{rr} = && 1 + \epsilon + \epsilon^2 + \epsilon^3,\\
g_T = && \left(1 + \frac{M}{r}\right)^2,\\
\tilde{\alpha} = && \left(1 + \frac{M}{r}\right)^{-1} \left(1 
+ \epsilon^2\right)^{-1} \left(1 + \frac{3M}{r}\right)^{-1},\\
\beta^r = && \epsilon^2 \left(1 + \frac{M}{r} \right) \left(1 
+ \epsilon^2\right)^{-1} \left(1 +\frac{3M}{r}\right)^{-1},\\
\hline \nonumber \\ 
K_{rr} = && - \frac{K_T}{g_T} \left(2 + \frac{3 \epsilon}{2} + \epsilon^2
+ \frac{\epsilon^3}{2}\right),\\
K_T = && \frac{4M^2}{r^3} \sqrt{\tilde{\alpha}},\\
f_{rrr} = && \frac{1}{g_T} \left[\frac{4}{r} (1 + \epsilon^2) + \frac{M}{r^2}
\left(11 - 2\epsilon + 9\epsilon^2\right)\right],\\
f_{rT} = && \frac{1}{r} + \frac{M}{r^2},
\end{eqnarray}
where
\begin{equation}
\epsilon \equiv \frac{2M}{r} \left(1 + \frac{M}{r}\right)^{-1}.
\end{equation}
\end{mathletters}
Here the horizon is located at $r=M$.


\subsection{Einstein-Klein-Gordon system}
\label{sec:Einst-Klein-Gord}

To add dynamics to the spherically symmetric problem, we introduce a
Klein-Gordon scalar field $\phi$ with stress-energy
\begin{equation}
\label{eq:KleinGordonStressEnergy}
4 \pi T_{ab} \equiv (\partial_a \phi)(\partial_b \phi)
- \frac{1}{2} g_{ab} (\partial_c \phi)(\partial^c \phi).
\end{equation}
Defining the quantities 
\begin{mathletters}
\begin{eqnarray}
\Pi \equiv && -N^{-1} \widehat{\partial_0} \phi,\\
\Phi_i \equiv && \partial_i \phi,
\end{eqnarray}
\end{mathletters}
the matter terms~(\ref{eq:MatterTermDefinitions}) are given by
\begin{mathletters}
\begin{eqnarray}
4 \pi \rho = && \frac{1}{2} \left(\Pi^2 + \Phi^i \Phi_i \right),\\
4 \pi J_i = && \Pi \Phi_i,\\
4 \pi T = && \Pi^2 - \Phi^i \Phi_i,\\
4 \pi S_{ij} = && \Phi_i \Phi_j + \frac{1}{2} g_{ij} (\Pi^2 - \Phi^i \Phi_i).
\end{eqnarray}
\end{mathletters}

The scalar field obeys $\Box \phi = 0$, which in spherical symmetry
takes the form
\end{multicols}
\begin{mathletters}
\label{KleinGordonWaveEq}
\begin{eqnarray}
\partial_t \Pi - \beta^r \partial_r \Pi + \frac{N}{g_{rr}} \partial_r \Phi_r 
= &&  N \left[ \Pi \left(\frac{K_{rr}}{g_{rr}} + \frac{2 K_T}{g_T} \right) 
- \frac{\Phi_r}{g_{rr}} \left(\frac{4 f_{rT}}{g_T} - \frac{2}{r}
+ \partial_r \ln \tilde{\alpha} \right)\right] ,\\
\partial_t \Phi_r - \beta^r \partial_r \Phi_r + N \partial_r \Pi 
= && - N \Pi \left(\frac{f_{rrr}}{g_{rr}} - \frac{2 f_{rT}}{g_T} 
- \frac{2}{r} + \partial_r \ln \tilde{\alpha} \right) 
+ \Phi_r \partial_r \beta^r.
\end{eqnarray}
\end{mathletters}
Only the derivatives of the scalar field $\phi$ appear in
equations~(\ref{eq:KleinGordonStressEnergy})
and~(\ref{KleinGordonWaveEq}), so $\phi$ itself need not be evolved.

For the evolution equations~(\ref{KleinGordonWaveEq}), the
characteristic fields in the radial direction ($\xi_r =
\sqrt{g_{rr}}$) are
\begin{equation}
U^\pm_\phi \equiv \Pi \pm \frac{\Phi_r}{\sqrt{g_{rr}}},
\end{equation}
with characteristic speeds $-\beta^r \pm \tilde{\alpha} g_T$.

\begin{multicols}{2}

\subsection{Apparent horizon}
A marginal outer trapped 2-surface is defined by the equation
\begin{equation}
\label{eq:MOTS}
D_i s^i + \left(s^i s^j -g^{ij}\right) K_{ij} = 0,
\end{equation}
where $D_i$ is the covariant derivative compatible with the
three-metric, and $s^i$ is the outward-pointing spatial unit normal of
the surface.  In spherical symmetry, equation~(\ref{eq:MOTS}) reduces
to
\begin{equation}
\label{eq:MOTS_spherical}
\frac{f_{rT}}{\sqrt{g_{rr}}} - K_T = 0.
\end{equation}
The apparent horizon is the outermost surface at which
(\ref{eq:MOTS_spherical}) is satisfied. On each time slice, the
coordinate radius of the horizon $r_{\text{ah}}$ is located by
solving~(\ref{eq:MOTS_spherical}) using a standard root-finding
algorithm.

If the horizon is to remain at a fixed coordinate radius as the
spacetime evolves, the following relation must be obeyed at the
horizon:
\begin{equation}
\label{eq:1DEC_freezehorizon}
\frac{\beta^r}{\tilde{\alpha} g_T} = \frac{1 + 8 \pi r^2 g_T (\rho - 
J_r/\sqrt{g_{rr}})}{1 - 8 \pi r^2 g_T(\rho - J_r/\sqrt{g_{rr}})}.
\end{equation}
Equation~(\ref{eq:1DEC_freezehorizon}) can be derived by setting the
time derivative of~(\ref{eq:MOTS_spherical}) equal to zero and
substituting the evolution and constraint equations to eliminate time
and spatial derivatives.


\subsection{Gauge conditions}

The question of which gauge conditions one should impose on a
numerically generated spacetime is one of the key unsolved problems in
numerical relativity.  In principle, the coordinate invariance of
general relativity allows one to make this choice
arbitrarily. However, a poor choice may not only obscure the physics
one is searching for in the simulation, but may also allow
rapidly-growing gauge modes that halt the code altogether.
\\ \mbox{} \hrulefill \mbox{} \\

\subsubsection{Algebraic conditions}

The simplest gauge choices we consider here are algebraic ones: we set
$\tilde{\alpha}$ and $\beta^i$ equal to their analytic values for some
parameterization of the Schwarzschild solution in
section~\ref{sec:Time-indep-slic} that we wish to reproduce
numerically. While these gauge conditions are obviously applicable
only to test problems, they provide a simplified setting in which to
study the properties of our evolution scheme. Such conditions have
been used extensively in 3D test problems\cite{bbhprl98a}.

Next we consider other algebraic gauge conditions that are independent
of a particular analytic solution, but are still of limited
generality.  For instance, one might require that the radial
coordinate remains areal, or in other words, that $g_T$ is
time-independent. Using our variables, this condition can be written
\begin{equation}
\label{eq:arealradiuscondition}
\beta^r f_{rT} - \tilde{\alpha} g_T \sqrt{g_{rr}} K_T = 0.
\end{equation}
One might also require that the ingoing coordinate speed of light
takes on a prescribed value $c_{-}$:
\begin{equation}
\label{eq:cminuscondition}
\tilde{\alpha} g_T + \beta^r + c_{-} = 0.
\end{equation}
These conditions are not generalizable to two black holes because the
first relies on the notion of an areal radial coordinate and the
second assumes that a unique ``ingoing'' direction exists at every
point in spacetime.  Nevertheless, these and similar gauge conditions
have proven useful for studies of single-black-hole spacetimes
\cite{marsa96,anninos_etal95}. Furthermore, if imposed only at one
point, they can be used as boundary conditions on more general
elliptic gauge choices, described below.

\subsubsection{Elliptic conditions}

We also explore gauge conditions that should be applicable to general
spacetimes.  For the shift vector, we consider two elliptic equations
due to \cite{smarryork78}: minimal strain
\end{multicols}
\begin{eqnarray}\label{eq:MinStrain1D}
\partial^2_r \beta^r && + \left(\frac{f_{rrr}}{g_{rr}} - \frac{2f_{rT}}{g_T}
\right) \partial_r \beta^r + \left[ \frac{\partial_r f_{rrr}}{g_{rr}}
- \frac{4 \partial_r f_{rT}}{g_T} 
- 2 \left(\frac{f_{rrr}}{g_{rr}}\right)^2
+ \frac{2 f_{rT}}{g_T} \left(\frac{5 f_{rrr}}{g_{rr}} - \frac{f_{rT}}{g_T}
- \frac{4}{r}\right)\right] \beta^r \nonumber \\ && \mbox{}
- \frac{K_{rr} g_T}{\sqrt{g_{rr}}} \partial_r \tilde{\alpha}
+ \tilde{\alpha} \sqrt{g_{rr}} g_T \left[ -\frac{\partial_r K_{rr}}{g_{rr}}
+ \frac{2 K_T f_{rT}}{(g_T)^2} 
+ \frac{K_{rr}}{g_{rr}} \left(\frac{f_{rrr}}{g_{rr}} + \frac{2}{r}
- \frac{8 f_{rT}}{g_T}\right) \right] = 0,
\end{eqnarray}
which minimizes changes in the three-metric in a global sense, and
minimal distortion,
\begin{eqnarray}\label{eq:MinDistortion1D}
\partial^2_r \beta^r && + \left(\frac{f_{rrr}}{g_{rr}} - \frac{2f_{rT}}{g_T}
\right) \partial_r \beta^r + \left[ \frac{\partial_r f_{rrr}}{g_{rr}}
- \frac{5 \partial_r f_{rT}}{g_T} 
- 2 \left(\frac{f_{rrr}}{g_{rr}}\right)^2
+ \frac{f_{rT}}{g_T} \left(\frac{11 f_{rrr}}{g_{rr}} - \frac{5 f_{rT}}{g_T}
- \frac{10}{r}\right)\right] \beta^r \nonumber \\ && \mbox{}
+ \left(\frac{K_T}{g_T} - \frac{K_{rr}}{g_{rr}}\right) \sqrt{g_{rr}} g_T 
\partial_r \tilde{\alpha} + \tilde{\alpha} \sqrt{g_{rr}} g_T \left[ 
-\frac{\partial_r K_{rr}}{g_{rr}} 
+ \frac{K_T}{g_T} \left( \frac{f_{rrr}}{g_{rr}} - \frac{2}{r} \right)
+ \frac{K_{rr}}{g_{rr}} \left(\frac{f_{rrr}}{g_{rr}} + \frac{2}{r}
- \frac{8 f_{rT}}{g_T}\right) \right] = 0,
\end{eqnarray}
which similarly minimizes changes of the conformal three-metric.

For the densitized lapse function, we consider the stationary mean
curvature condition $\partial_t K=0$, which in terms of our variables
can be written
\begin{eqnarray}\label{eq:StatMeanCurv1D}
\partial^2_r \tilde{\alpha} && +  \left( \frac{f_{rrr}}{g_{rr}}
+ \frac{2 f_{rT}}{g_T} - \frac{4}{r}\right) \partial_r \tilde{\alpha}
+ \tilde{\alpha} \left[ \frac{\partial_r f_{rrr}}{g_{rr}} 
+ \frac{2 \partial_r f_{rT}}{g_T} 
- 2 \left( \frac{f_{rrr}}{g_{rr}} + \frac{f_{rT}}{g_T} \right)
\left( \frac{f_{rrr}}{g_{rr}} - \frac{5 f_{rT}}{g_T} \right) + \frac{6}{r^2}
\right. \nonumber \\ && \left. \mbox{} - \frac{2 g_{rr}}{r^2 g_T} 
- g_{rr} \left( \frac{K_{rr}}{g_{rr}} + \frac{2 K_T}{g_T}\right)^2
- \frac{2 f_{rrr}}{r g_{rr}} +8 \pi S_{rr}
- 4 \pi g_{rr} T \right]  \nonumber \\ && \mbox{}
- \frac{\beta^r \sqrt{g_{rr}}}{g_T} \left[ \frac{\partial_r K_{rr}}{g_{rr}} 
+ \frac{2 \partial_r K_T}{g_T} + \frac{4 K_T}{g_T} \left( \frac{1}{r}
- \frac{f_{rT}}{g_T} \right) 
+ \frac{2 K_{rr}}{g_{rr}} \left( \frac{4 f_{rT}}{g_T} 
- \frac{f_{rrr}}{g_{rr}} \right)  \right] = 0.
\end{eqnarray}
\begin{multicols}{2}
This condition was also discussed by \cite{smarryork78}, and is best
known in the special case $K=0$, when it reduces to the familiar
maximal slicing condition. Use of the stationary mean curvature lapse
combined with either the minimal strain or minimal distortion shift
vectors has been recently encouraged by
\cite{Garfinkle2000,Garfinkle1999}.

Each of the elliptic
equations~(\ref{eq:MinStrain1D}),~(\ref{eq:MinDistortion1D}),
and~(\ref{eq:StatMeanCurv1D}) requires two boundary conditions.  At
the horizon, we either set $\tilde\alpha$ and $\beta^r$ to prescribed
values, or we impose~(\ref{eq:1DEC_freezehorizon})
and~(\ref{eq:cminuscondition}).  At the outer boundary, we again can
set $\tilde\alpha$ and $\beta^r$ to prescribed values, we can
impose~(\ref{eq:arealradiuscondition}) and~(\ref{eq:cminuscondition}),
or we can impose Robin conditions of the form~(\ref{eq:Robin}) on
$\tilde\alpha$ and $\beta^r$.


\subsection{Mass}

It is useful for diagnostic purposes to compute the mass of the
spacetime.  In spherical symmetry, the total mass inside an invariant
spherical surface labeled by coordinate $r$ is well-defined and given
by the Misner-Sharp formula \cite{MTW}, which for our variables reads
\begin{equation}
M_{\text{MS}}(r) \equiv \frac{r\sqrt{g_T}}{2} \left[ 1 +
\frac{r^2}{g_T} \left(K_T^2 - \frac{f_{rT}^2}{g_{rr}} \right) \right].
\end{equation}

\section{Pseudospectral collocation methods}
\label{sec:Pseud-coll-meth}

\subsection{Introduction}
Consider a system of $L$ evolution equations of the form
\begin{equation}
\label{eq:gensystem}
\partial_t f^{(\ell)} = {\cal F}^{(\ell)} [\{f^{(\ell)}\}]
\end{equation}
for $1 \leq \ell \leq L$, where $\{f^{(\ell)}(\vec x,t)\}$ is the
solution, and ${\cal F}^{(\ell)} [\{f^{(\ell)}\}]$ are (possibly
nonlinear) functions of $\{f^{(\ell)}\}$ and their spatial
derivatives.  Approximate each function $f^{(\ell)}$ of the solution
as a finite sum of basis functions $\phi_k^{(\ell)}(\vec x)$,
\begin{equation}
f^{(\ell)}_N (\vec x,t) = \sum_{k=0}^{N-1} \tilde f_k^{(\ell)} (t)
\phi_k^{(\ell)} (\vec x).
\end{equation}
For smooth functions as $N\to\infty$ the approximation is exact.
Corresponding to the approximate solution $\{f^{(\ell)}_N\}$ is a
residual
\begin{equation}
R^{(\ell)}_N = \partial_t f^{(\ell)}_N - {\cal F}^{(\ell)}
[\{f^{(\ell)}_N\}],
\end{equation}
for each evolution equation.

In PSC the spectral coefficients $\tilde f_k^{(\ell)} (t)$ are
determined by demanding that the residuals $R^{(\ell)}_N$ vanish at a
fixed set of $N$ collocation points $\vec x_n$.  In other words, it is
demanded that the system of differential
equations~(\ref{eq:gensystem}) be satisfied exactly at the collocation
points $\{\vec x_n \}$.  The choice of the collocation points is
intimately related to the choice of basis functions used in the
approximate solution.  In the following subsection, we discuss how
they are chosen.


\subsection{Expansion basis and collocation points}

For the remainder of this section, we will restrict ourselves to
problems with one spatial dimension (1D).  The choice of an expansion
basis depends upon the particular problem being solved.  For example,
the natural expansion basis for a 1D problem with periodic boundary
conditions is a Fourier series. For more general boundary conditions,
such as the ones we will impose in our black hole evolutions,
Chebyshev polynomials are a robust choice for the basis functions.
Chebyshev polynomials are defined on the interval
\begin{equation}
{\Bbb{I}} = [-1,1]
\end{equation}
by
\begin{equation}
T_k (x) = \cos ( k \cos^{-1} x) .
\end{equation}
\end{multicols}
\twocolumn \noindent
A function $f$ on $\Bbb{I}$ is approximated as \footnote{For Chebyshev
bases the conventional notation is that $k$ runs from $0$ to $N$, not
$N-1$; thus, there are $N+1$ coefficients and collocation points.}
\begin{equation}
f_N (x,t) = \sum_{k=0}^{N} \tilde f_k (t) T_k (x).
\label{eq:chebexpansion}
\end{equation}
Note that in order to use this expansion, we must specify a mapping
from our physical domain $[r_{\text{min}},r_{\text{max}}]$ to
$\Bbb{I}$.  The simplest choice is a linear mapping, but other choices
may work better.

For a Chebyshev expansion, a convenient choice of the collocation
points is
\begin{equation}
\label{eq:Chebyshev_Collocation_Points}
x_n = \cos \frac{\pi n}{N} .
\end{equation}
At these collocation points, the Chebyshev polynomials satisfy the
discrete orthogonality relation
\begin{equation}
\delta_{jk} = \frac{2}{N \bar{c}_k} \sum_{n = 0}^{N} \frac{1}{\bar{c}_{n}} 
T_{j}(x_{n}) T_{k}(x_{n}),
\end{equation}
where
\begin{equation}
\bar{c}_k = \left\{ 
\begin{array}{ll}
2, & \; \text{$k = 0$ or $N$} \\
1, & \; \text{$1 \leq k \leq N-1$}.
\end{array}
\right.
\end{equation}
Using the orthogonality relation, the spectral coefficients are given
by
\begin{equation}
\label{eq:speccoefs}
\tilde f_k = \frac{2}{N \bar{c}_k} \sum_{n = 0}^{N} \frac{1}{\bar{c}_{n}} 
f_N(x_{n}) T_{k}(x_{n}).  
\end{equation}
Since
\begin{equation}
T_k (x_n) = \cos \frac{\pi k n}{N},
\end{equation}
fast cosine transforms can be used to compute~(\ref{eq:chebexpansion})
at the collocation points and to evaluate~(\ref{eq:speccoefs}).

In PSC, the focus is not on the set of spectral coefficients $\{\tilde
f_k (t)\}$, but on the equivalent set $\{f (x_n,t)\}$, the approximate
solution evaluated at the collocation points.  In particular, the
approximate solution to~(\ref{eq:gensystem}) would be given by
evolving
\begin{equation}
\label{eq:approxsystem}
\partial_t f^{(\ell)}_N(x_n,t) = {\cal F}^{(\ell)}(x_n,t),
\end{equation}
for $1 \leq \ell \leq L$ and $0 \leq n \leq N$.  Given initial
conditions $f^{(\ell)}(x,0)$, and appropriate boundary conditions,
Equation~(\ref{eq:approxsystem}) can be evolved forward in time using
the method of lines, described in section~\ref{sec:Appl-bound-cond}.
Since the focus is on grid-point values, and not the spectral
coefficients, it is possible to reuse large amounts of code developed
for FD methods.


\subsection{Computation of derivatives}

The main differences between PSC and FD in
evolving~(\ref{eq:approxsystem}) are the choice of collocation (grid)
points $x_n$, how spatial derivatives are computed, and how boundary
conditions are imposed.  In PSC, spatial derivatives are computed
analytically from the series expansion
\begin{equation}
\frac {\partial f_N (x,t)}{\partial x} = \sum_{k=0}^{N} \tilde f_k (t)
\frac {d T_k (x)}{dx}.
\end{equation}
This derivative can be written as another sum over Chebyshev
polynomials
\begin{equation}
\label{eq:derivative}
\frac {\partial f_N (x,t)}{\partial x} = \sum_{k=0}^{N} {\tilde f}'_k (t) 
T_k (x),
\end{equation}
by using the simple recursion relation
\begin{equation}
\label{eq:recursion}
c_k {\tilde f_k}'(t) = {\tilde f_{k+2}}'(t) + 2 (k+1) \tilde f_{k+1}(t),
\end{equation}
where
\begin{equation}
{c}_k = \left\{ 
\begin{array}{ll}
2, & \; \text{$k = 0$} \\
1, & \; \text{$k \ge 1$}.
\end{array}
\right.
\end{equation}

Evaluating a derivative requires two fast transforms; the first to
compute the spectral coefficients needed in the recursion
relation~(\ref{eq:recursion}), the second to
evaluate~(\ref{eq:derivative}).


\subsection{Time evolution and application of boundary conditions}
\label{sec:Appl-bound-cond}

We evolve our hyperbolic system using the method of lines.  In this
method, we cast our system into the form~(\ref{eq:approxsystem}) and
use a standard ODE solver to integrate the equation in time.  For most
of the results presented in this paper, we have used a fourth-order
explicit Runge-Kutta method.  One of the drawbacks of using PSC is
that the Chebyshev collocation
points~(\ref{eq:Chebyshev_Collocation_Points}) are clustered near the
domain boundaries.  This places a more severe Courant stability limit
$\Delta t \sim O(N^{-2})$ on a wave equation than for FD, where
$\Delta t \sim \Delta x \sim O(N^{-1})$. Because of the superior
spatial convergence of PSC, however, this restriction is not as severe
as it may seem at first glance. In fact, to retain the accuracy gained
by the spatial resolution, it may be necessary to use a time step
smaller than that demanded by stability.  Moreover, if the stability
restriction on the time limit becomes too severe for practical
evolutions, one can implement implicit or semi-implicit time-stepping
schemes.

One of the advantages of PSC over FD is in how the boundary conditions
are applied.  In FD, derivatives are approximated by differences of
field variables at grid points. The pattern of grid points used must
typically be modified at the boundaries of the numerical grid.
Consequently, boundary conditions can be difficult to formulate in FD.
In PSC, on the other hand, the approximate solution is given over the
entire domain.  As seen in the previous section, derivatives are
computed analytically; therefore nothing special needs to be done to
compute the derivative at a boundary.  Furthermore, since there are
collocation points on the boundary of the domain, the application of
boundary conditions is straightforward in PSC.  One simply demands
that the approximate solution satisfy the exact boundary condition at
the boundary collocation point.

The boundary conditions are applied during the time step by modifying
${\cal F}^{(\ell)} [\{f^{(\ell)}\}]$ (cf. equation~\ref{eq:gensystem})
at the boundary points so that the boundary conditions are satisfied.
In this paper we are interested in applying boundary conditions on a
hyperbolic system of evolution equations.  As described in
section~\ref{sec:Boundary-conditions}, the solution to a hyperbolic
system can be written in terms of characteristic fields that propagate
with corresponding characteristic speeds.  Physically we know that
boundary conditions need only be applied to the incoming
characteristic fields.

Therefore, to impose a boundary condition at a domain boundary
$x=x_b$, we first compute the time derivatives of the characteristic
fields $U_c(x_b)$ at the boundary.  We then apply a boundary condition
to those fields that are propagating into the domain; the remaining
fields are untouched.  Finally, we reconstruct the time derivatives of
the fundamental variables at $x_b$ and use these values in the time
update.  Failure to impose a boundary condition on an incoming field
or imposition of a boundary condition on an outgoing field almost
always leads to an unstable evolution.

A Dirichlet boundary condition $u_c(x_b,t) = g(t)$ is applied by
ensuring that the time derivative of the incoming characteristic field
at the boundary collocation point is set to $dg/dt$.  A boundary
condition such as Neumann or Robin that involves the spatial
derivative of the characteristic field is enforced by replacing the
spatial derivative at the boundary with the appropriate value in order
to satisfy the boundary condition.
\subsection{Multiple domains}
In order to use a PSC method for problems of dimension $d$ greater
than unity the computational domain must be sufficiently simple that
it can be mapped to ${ \Bbb{I}}^{d}$ or ${ \Bbb{I}}^{d-2} \times S^2$
(where $S^2$ are two-spheres). For three dimensions, this typically
means a cube, a sphere, or a spherical shell.  If the computational
domain is more complicated, then it must be decomposed into
sub-domains that can each be mapped to one of these domains.  For
example, in two dimensions an L-shaped region can be decomposed into
two adjacent rectangles.

The binary black hole problem will need to be solved using
multiple domains.  Therefore we test our ability to handle multiple
domains on our one-dimensional problems.  The use of multiple domains
also provides a natural way of making our code run in parallel.  We
use KeLP \cite{KeLP} to handle communication between multiple domains
and for parallelization of our code.

The extension of our method from one domain to multiple domains is
straightforward.  We evolve each domain independently with
communication done only at the boundaries.  At the domain boundaries
we compute the time derivatives of the characteristic fields in each
domain.  We then replace the time derivatives of the incoming
characteristic fields at the boundary with the time derivatives of the
outgoing characteristic fields of the neighboring domain.  If there is
no neighboring domain at a particular boundary, the external boundary
condition is applied as described in
section~\ref{sec:Appl-bound-cond}.


\subsection{Solving elliptic equations}

In addition to evolving our hyperbolic system of evolution
equations~(\ref{eq:3DEC}), we may need to solve elliptic equations in
order to construct initial data for the Einstein-Klein-Gordon system
or to enforce elliptic gauge conditions.  Consider a linear elliptic
equation of the form
\begin{equation}
{\cal L} (u(x)) = f(x),
\end{equation}
where $u(x)$ is the solution we are seeking.  This can be cast as a
matrix problem where, unlike for FD, the matrix corresponding to the
linear operator ${\cal L}$ is full.  In 1D we solve this matrix
equation directly, but for higher-dimensional problems, it will be
more efficient to use an iterative method.

A nonlinear elliptic equation such as the Hamiltonian constraint can
be solved either by the methods described in \cite{Kidder2000}, or by
linearizing the nonlinear system and iterating the linearized
equations until a solution is found.  The latter method is employed in
the work described here.


\subsection{Filtering}

The errors in a spectral method are dominated by two types of terms of
roughly equal magnitude.  Truncation error arises from the neglect of
the high-frequency terms that are not retained in the truncated
series.  Aliasing error occurs because each neglected high-frequency
mode is indistinguishable from some retained lower-frequency mode when
sampled only at the collocation points; for example, the functions
$\sin(\pi x/5)$ and $\sin(-9\pi x/5)$ take the same values on a grid
of $N$ points $x\in\{0,1,2,\ldots,N-1\}$. Because of aliasing error,
power in the high-frequency mode, instead of being completely
neglected, ends up contributing to the lower-frequency mode.

When solving a nonlinear system of equations it becomes important to
control the aliasing error.  This can be done by filtering the
high-frequency modes of the retained series.  For quadratic
nonlinearities it is sufficient to zero the top third of the spectral
coefficients to eliminate aliasing \cite{Orszag1971}. In our 1D
evolutions, we have found it necessary to filter only gauge variables
that are computed from an elliptic equation.  In effect we are
smoothing the solutions to the elliptic gauge equations to eliminate
high-frequency noise.  Our preliminary investigations suggest
that more extensive filtering may be required to produce stable
evolutions in 3D.


\section{Numerical results}
\label{sec:Numerical-results}

\subsection{Schwarzschild black hole}

In this section we evolve a time-independent slicing of a
Schwarzschild black hole.  We begin our numerical evolutions with
initial data corresponding to one of the slicings given in
Sec.~\ref{sec:Time-indep-slic}.  If the evolution equations are
integrated exactly, the solution will remain time-independent.  We can
test the convergence of our method by measuring the deviation of the
solution from the initial data at a given coordinate time, or by
measuring the constraint quantities~(\ref{eq:1DEC:constraints}), which
are zero for the exact solution. For all evolutions, the interior of
the hole is excised, and no boundary condition is applied at the inner
boundary because all characteristic fields are outgoing (off the
domain) there. In Table~\ref{tableofruns} we list the input parameters
and the results for selected evolutions.

\begin{table}
\caption{Input parameters for selected evolutions.  For each evolution
we list the initial data type, the locations $r/M$ of the inner and
outer boundaries, the outer boundary conditions on the incoming fields
$U^-_T$ and on $\{U^0_r,U^0_t\}$, the gauge
conditions (including boundary conditions for elliptic equations),
the time stepping algorithm, and the result of the evolution.}
\label{tableofruns}
\begin{tabular}{rlddllllll}
Run&ID\tablenote{PG: Painlev\'{e}-Gullstrand; KS: Kerr-Schild;
H: harmonic time; FH: fully harmonic}&
IB&OB&\multicolumn{2}{c}{BC\tablenote{F: freezing; C: constraint}}&
\multicolumn{1}{c}{Lapse\tablenote{C: constant;
S({\it ib})({\it ob}): stationary mean curvature; F: freeze;
c{\it n}: $c_- = -{\it n}$ at horizon or outer boundary}}&
\multicolumn{1}{c}{Shift\tablenote{C: constant;
MS({\it ib})({\it ob}): minimal strain;
MD({\it ib})({\it ob}): minimal distortion;
X: equation~(\ref{eq:1DEC_freezehorizon});
R: Robin; A: $\partial_t g_T=0$}}&
TS\tablenote{R4: 4th-order Runge-Kutta; BE: backward Euler}&
\multicolumn{1}{c}{Res\tablenote{Stb: stable; Exp: exponential growth; 
LG: linearly-growing gauge mode;
QG: quadratically-growing gauge mode}}\\
\hline
\refstepcounter{run}\label{LKWF03-30-2000}\ref{LKWF03-30-2000}
&KS&1.9&11.9&F&F&C&C&R4&Stb\\
\refstepcounter{run}\label{O04-25-2000B}\ref{O04-25-2000B}
&KS&1.9&11.9&C&F&C&C&R4&Stb\\
\refstepcounter{run}\label{O04-25-2000A}\ref{O04-25-2000A}
&KS&1.9&11.9&F&C&C&C&R4&Stb\\
\refstepcounter{run}\label{O04-25-2000C}\ref{O04-25-2000C}
&KS&1.9&11.9&C&C&C&C&R4&Stb\\
\refstepcounter{run}\label{LKO05-10-2000}\ref{LKO05-10-2000}
&KS&1.75&120&C&C&C&C&R4&Stb\\
\refstepcounter{run}\label{WF03-30-2000}\ref{WF03-30-2000}
&KS&1.9&11.9&F&F&S(c1)(c1)&MD(X)(A)&R4&Exp\\
\refstepcounter{run}\label{WF04-24-2000C}\ref{WF04-24-2000C}
&KS&1.9&11.9&F&F&S(c1)(c1)&MS(X)(A)&R4&Exp\\
\refstepcounter{run}\label{WF04-24-2000B}\ref{WF04-24-2000B}
&KS&1.9&11.9&C&F&S(c1)(c1)&MS(X)(A)&R4&LG\\
\refstepcounter{run}\label{WF04-24-2000A}\ref{WF04-24-2000A}
&KS&1.9&11.9&F&C&S(c1)(c1)&MS(X)(A)&R4&Exp\\
\refstepcounter{run}\label{WF04-18-2000H}\ref{WF04-18-2000H}
&KS&1.9&11.9&C&C&S(c1)(c1)&MS(X)(A)&R4&LG\\
\refstepcounter{run}\label{WF04-24-2000D}\ref{WF04-24-2000D}
&KS&1.9&11.9&C&C&S(c1)(c1)&MD(X)(A)&R4&LG\\
\refstepcounter{run}\label{WF04-17-2000B}\ref{WF04-17-2000B}
&KS&1.9&11.9&C&C&S(c1)(c1)&MS(X)(A)&BE&LG\\
\refstepcounter{run}\label{LKWF03-30-2000A}\ref{LKWF03-30-2000A}
&PG&1.9&11.9&F&F&C&C&R4&Stb\\
\refstepcounter{run}\label{O04-24-2000B}\ref{O04-24-2000B}
&PG&1.9&11.9&C&F&C&C&R4&Stb\\
\refstepcounter{run}\label{O04-24-2000A}\ref{O04-24-2000A}
&PG&1.9&11.9&F&C&C&C&R4&LG\\
\refstepcounter{run}\label{O04-24-2000C}\ref{O04-24-2000C}
&PG&1.9&11.9&C&C&C&C&R4&LG\\
\refstepcounter{run}\label{LKO05-10-2000E}\ref{LKO05-10-2000E}
&PG&1.75&120&C&C&C&C&R4&Stb\\
\refstepcounter{run}\label{LKWF04-20-2000B}\ref{LKWF04-20-2000B}
&PG&1.9&11.9&F&F&S(c2)(F)&MS(X)(R)&R4&Exp\\
\refstepcounter{run}\label{LKWF04-20-2000D}\ref{LKWF04-20-2000D}
&PG&1.9&11.9&C&F&S(c2)(F)&MS(X)(R)&R4&Exp\\
\refstepcounter{run}\label{LKWF04-20-2000C}\ref{LKWF04-20-2000C}
&PG&1.9&11.9&F&C&S(c2)(F)&MS(X)(R)&R4&LG\\
\refstepcounter{run}\label{LKWF04-20-2000A}\ref{LKWF04-20-2000A}
&PG&1.9&11.9&C&C&S(c2)(F)&MS(X)(R)&R4&LG\\
\refstepcounter{run}\label{LKWF04-24-2000}\ref{LKWF04-24-2000}
&PG&1.9&11.9&C&C&S(c2)(F)&MS(X)(R)&BE&LG\\
\refstepcounter{run}\label{WF04-18-2000G}\ref{WF04-18-2000G}
&H&1.9&11.9&F&F&C&C&R4&Exp\\
\refstepcounter{run}\label{WF04-10-2000}\ref{WF04-10-2000}
&H&1.9&3.9&F&F&C&C&R4&Stb\\
\refstepcounter{run}\label{O04-25-2000F}\ref{O04-25-2000F}
&H&1.9&11.9&C&F&C&C&R4&Exp\\
\refstepcounter{run}\label{O04-25-2000E}\ref{O04-25-2000E}
&H&1.9&11.9&F&C&C&C&R4&Exp\\
\refstepcounter{run}\label{WF04-17-2000D}\ref{WF04-17-2000D}
&H&1.9&11.9&C&C&C&C&R4&QG\\
\refstepcounter{run}\label{WF04-14-2000B}\ref{WF04-14-2000B}
&H&1.9&11.9&C&C&C&C&BE&QG\\
\refstepcounter{run}\label{LKO05-10-2000C}\ref{LKO05-10-2000C}
&H&1.75&120&C&C&C&C&R4&LG\\
\refstepcounter{run}\label{WF04-18-2000E}\ref{WF04-18-2000E}
&H&1.9&11.9&C&C&S(c$\frac{1}{2}$)(F)&MS(X)(A)&BE&LG\\
\refstepcounter{run}\label{WF04-18-2000I}\ref{WF04-18-2000I}
&H&1.9&11.9&C&C&S(c$\frac{1}{2}$)(F)&MS(X)(A)&R4&LG\\
\refstepcounter{run}\label{O04-26-2000}\ref{O04-26-2000}
&FH&0.9&10.9&F&F&C&C&R4&Exp\\
\refstepcounter{run}\label{O04-26-2000A}\ref{O04-26-2000A}
&FH&0.9&10.9&F&C&C&C&R4&Exp\\
\refstepcounter{run}\label{O04-26-2000B}\ref{O04-26-2000B}
&FH&0.9&10.9&C&F&C&C&R4&Exp\\
\refstepcounter{run}\label{LKWF05-12-2000A}\ref{LKWF05-12-2000A}
&FH&0.9&10.9&C&C&C&C&R4&QG\\
\refstepcounter{run}\label{LKO05-12-2000B}\ref{LKO05-12-2000B}
&FH&0.9&120&C&C&C&C&R4&LG\\
\refstepcounter{run}\label{O04-26-2000H}\ref{O04-26-2000H}
&FH&0.9&6.9&F&F&C&C&BE&Stb\\
\refstepcounter{run}\label{O04-26-2000K}\ref{O04-26-2000K}
&FH&0.9&7.9&F&F&C&C&BE&Exp\\
\refstepcounter{run}\label{WF05-02-2000A}\ref{WF05-02-2000A}
&FH&0.9&10.9&C&C&S(c$\frac{1}{2}$)(F)&MS(X)(A)&BE&LG\\
\end{tabular}
\end{table}


\subsubsection{Analytic gauge conditions}
\label{sec:Analyt-gauge-cond}

\begin{figure}
\begin{center}
\begin{picture}(240,240)
\put(0,0){\epsfxsize=3.5in\epsffile{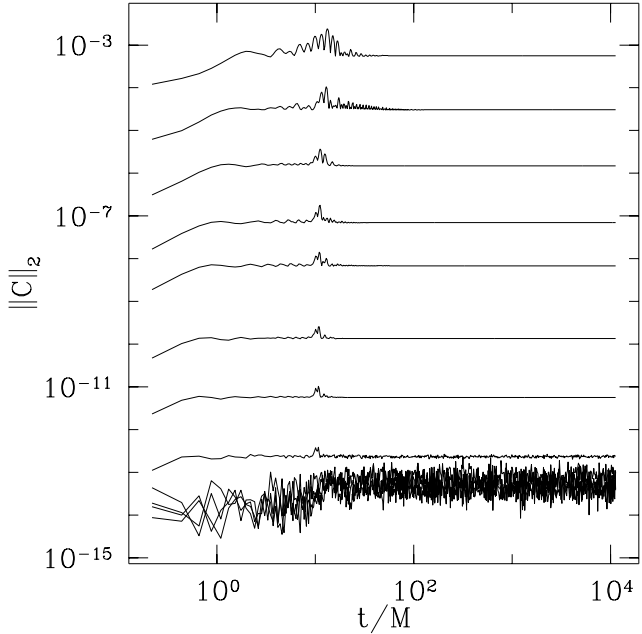}}
\end{picture}
\end{center}
\caption{Long-term stability of the evolution of Kerr-Schild initial
data, run~\ref{LKWF03-30-2000} from Table~\ref{tableofruns}.  Plotted
is the $\ell_2$ norm of the Hamiltonian constraint~(\ref{eq:hamcon})
in units of $M^{-2}$ as a function of time for several spatial
resolutions. The number of spectral coefficients $N_r$ for each plot,
starting at the top, is $12$, $16$, $20$, $24$, $27$, $32$, $36$,
$40$, $45$, $48$, $54$, and $60$.}
\label{fig:ksanlygauge_stab}
\end{figure}

\begin{figure}
\begin{center}
\begin{picture}(240,240)
\put(0,0){\epsfxsize=3.5in\epsffile{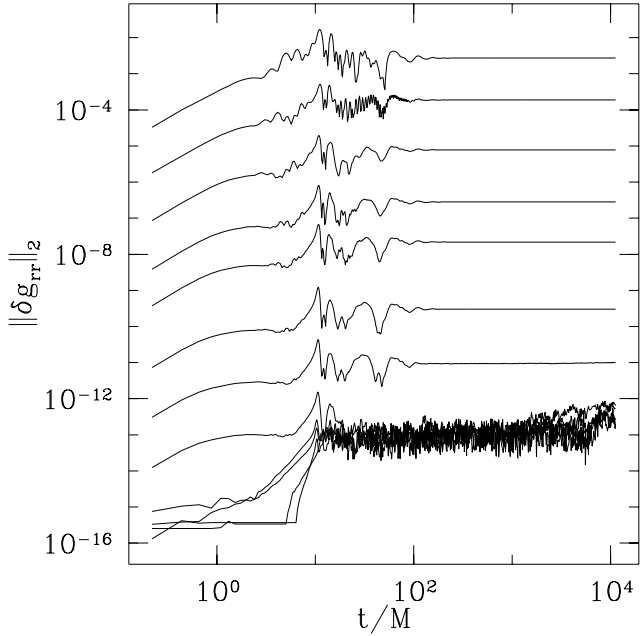}}
\end{picture}
\end{center}
\caption{Norm of the error in $g_{rr}$ as a function of time for the
same evolutions shown in Fig.~\ref{fig:ksanlygauge_stab}.}
\label{fig:ksanlygaugedeltagrr_stab}
\end{figure}

\begin{figure}
\begin{center}
\begin{picture}(240,240)
\put(0,0){\epsfxsize=3.5in\epsffile{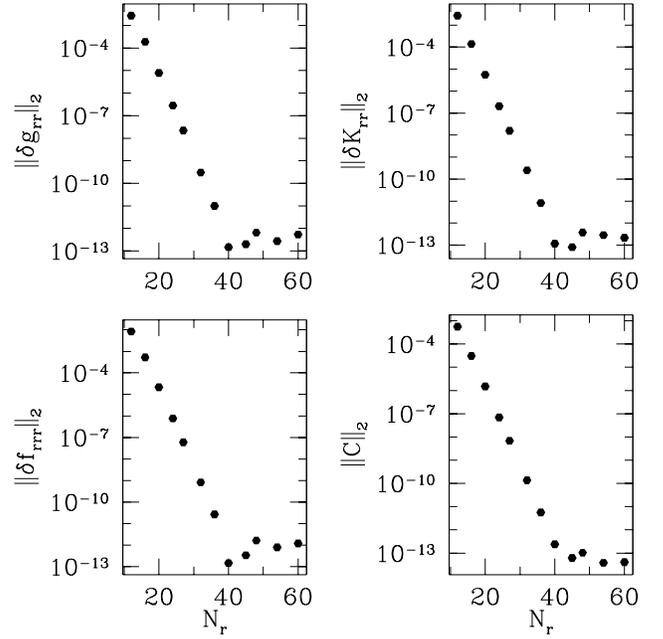}}
\end{picture}
\end{center}
\caption{Norms of the Hamiltonian constraint and errors in selected
fundamental variables plotted as a function of the number of spectral
coefficients $N_r$ at $t = 11000 M$ for the evolutions shown in
Fig.~\ref{fig:ksanlygauge_stab}. The quantities $\delta K_{rr}$ and
$\delta f_{rrr}$ are measured in units of $M^{-1}$. The errors
decrease exponentially with $N_r$.}
\label{fig:ksanlygaugeConvergence}
\end{figure}

The simplest gauge treatment is to fix the gauge variables
$\tilde\alpha$ and $\beta^r$ to their initial values during the entire
evolution. For example, one can begin the evolution with Kerr-Schild
initial data~(\ref{kerr-schild}) and set $\tilde\alpha$ and $\beta^r$
according to the analytic
expressions~(\ref{kerr-schild-lapse}--\ref{kerr-schild-shift}) for all
time.  In Figs.~\ref{fig:ksanlygauge_stab}
and~\ref{fig:ksanlygaugedeltagrr_stab} we plot the norm of the
Hamiltonian constraint and the deviation of $g_{rr}$ from the analytic
solution versus time for such an evolution (run~\ref{LKWF03-30-2000}
from Table~\ref{tableofruns}).  Each plot shows results for several
spatial resolutions $N_r$ run at a fixed time resolution $\Delta t =
0.007M$. The features near $t = 10M$ correspond to a small error pulse
that begins at the outer boundary at $t=0$, grows like $r^{-2}$ as it
propagates inwards, and eventually falls into the hole. After several
crossing times, the evolution settles into a steady state that
converges to the analytic solution as one increases the spatial
resolution.  We end the evolutions at $t=11000M$ even though they
clearly would have proceeded further.  The convergence rate is
exponential until machine roundoff errors dominate, as illustrated in
Fig.~\ref{fig:ksanlygaugeConvergence}.  Repeating the evolutions
shown in
Figs.~\ref{fig:ksanlygauge_stab}--\ref{fig:ksanlygaugeConvergence}
for Painlev\'{e}-Gullstrand initial data (run~\ref{LKWF03-30-2000A}
from Table~\ref{tableofruns}) yields similar results.

\begin{figure}
\begin{center}
\begin{picture}(240,240)
\put(0,0){\epsfxsize=3.5in\epsffile{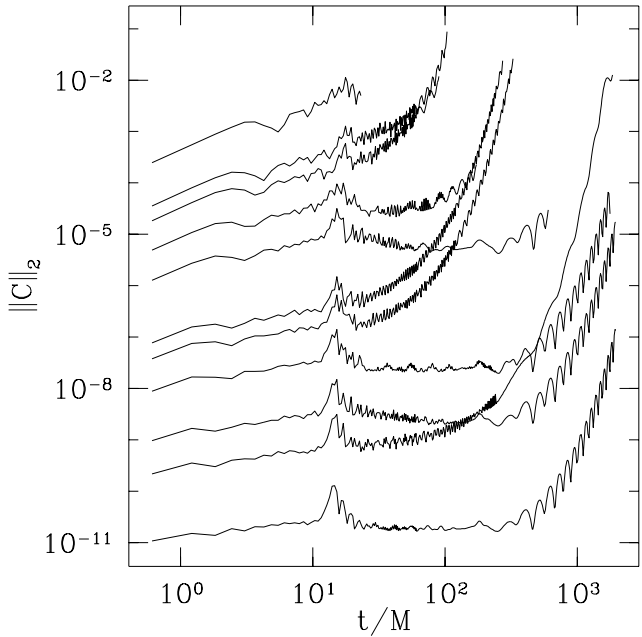}}
\end{picture}
\end{center}
\caption{Norm of the Hamiltonian constraint~(\ref{eq:hamcon}) as a
function of time for several spatial resolutions for evolutions of
harmonic initial data, run~\ref{WF04-18-2000G} of
Table~\ref{tableofruns}. The number of spectral coefficients $N_r$ for
each plot, starting at the top, is $12$, $15$, $16$, $18$, $20$, $24$,
$25$, $27$, $30$, $32$, and $36$.}
\label{fig:WF04-18-2000GECHl2}
\end{figure}

In Fig.~\ref{fig:WF04-18-2000GECHl2} we show the norm of the
Hamiltonian constraint for run~\ref{WF04-18-2000G} of
Table~\ref{tableofruns}. This evolution is identical to
run~\ref{LKWF03-30-2000} except the initial data, as well as the
values of $\tilde{\alpha}$ and $\beta^r$ for all time, correspond to a
time-independent harmonic slice of the Schwarzschild
geometry~(\ref{harmonic_time}).  Rather than settling to a steady
state, the numerical solution grows exponentially at late times,
eventually crashing the code. This is caused by a combination of
high-frequency numerical instabilities, rapidly-growing gauge modes,
and rapidly-growing constraint violating modes, all of which can be
suppressed by appropriate changes in the evolution algorithm, as
described below. Evolutions of fully harmonic initial
data~(\ref{full_harmonic}) behave similarly. It is not known why these
instabilities are absent in evolutions of Kerr-Schild and
Painlev\'{e}-Gullstrand initial data.  However, the dependence of
stability on the choice of initial data should not be too surprising
if one thinks of the initial data as a background solution and the
numerical evolution as a perturbation on this background: in general,
modifying the background solution can change the stability of
perturbations.

\begin{figure}
\begin{center}
\begin{picture}(240,240)
\put(0,0){\epsfxsize=3.5in\epsffile{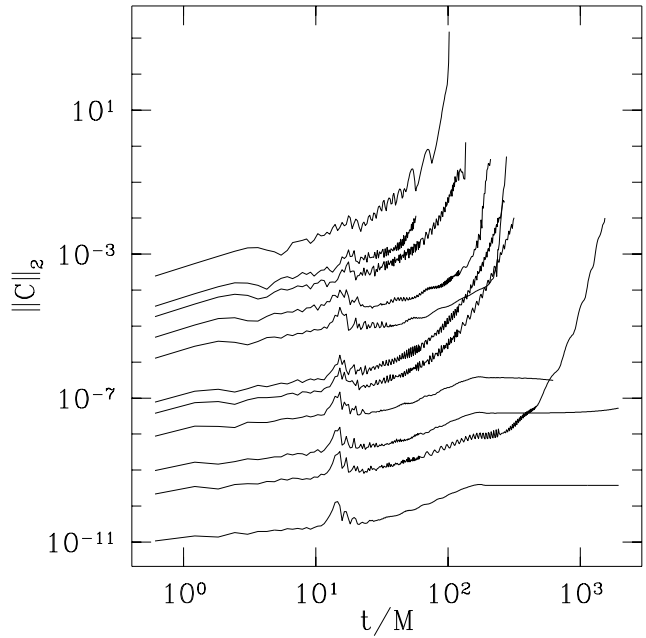}}
\end{picture}
\end{center}
\caption{Norm of the Hamiltonian constraint~(\ref{eq:hamcon}) as a
function of time for several spatial resolutions for evolutions of
harmonic initial data, run~\ref{WF04-17-2000D} of
Table~\ref{tableofruns}. Constraint-based outer boundary conditions
are imposed on $U^0_r$, $U^0_t$, and $U^-_T$.  Resolutions
are the same as in Fig.~\ref{fig:WF04-18-2000GECHl2}.}
\label{fig:WF04-17-2000DECHl2}
\end{figure}

\begin{figure}
\begin{center}
\begin{picture}(240,240)
\put(0,0){\epsfxsize=3.5in\epsffile{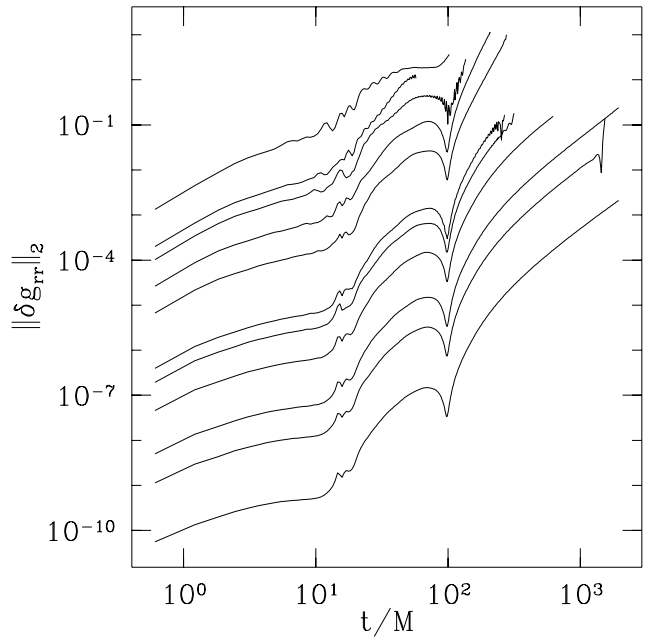}}
\end{picture}
\end{center}
\caption{Error in $g_{rr}$ versus time for the same evolutions shown
in Fig.~\ref{fig:WF04-17-2000DECHl2}. The growth is quadratic in $t$
at late times.}
\label{fig:WF04-17-2000Dgrrdiffl2}
\end{figure}

For the evolutions shown in Fig.~\ref{fig:WF04-18-2000GECHl2},
freezing boundary conditions~(\ref{eq:ConstantOuterBoundary}) are
imposed on the incoming characteristic fields $U^0_r$, $U^0_t$,
$U^-_r$, and $U^-_T$.  One can suppress the
constraint-violating modes seen in Fig.~\ref{fig:WF04-18-2000GECHl2}
by replacing the freezing boundary conditions on $U^0_r$, $U^0_t$, and
$U^-_T$ with constraint boundary conditions as discussed in
Section~\ref{sec:Boundary-conditions}.  The resulting evolutions are
shown in Figs.~\ref{fig:WF04-17-2000DECHl2}
and~\ref{fig:WF04-17-2000Dgrrdiffl2}. Except for the evolution with
$N_r=32$ discussed below, the Hamiltonian constraint ${\cal C}$
settles to a steady state that converges exponentially to zero.  The
same is true for the other three constraints ${\cal C}_{rT}$, ${\cal
C}_{rrr}$, and ${\cal C}_r$. However, the metric quantities and other
fundamental variables grow approximately quadratically with time,
eventually causing the simulations to terminate.  Because the
constraints remain satisfied, we attribute this quadratic growth to a
gauge mode.

\begin{figure}
\begin{center}
\begin{picture}(240,240)
\put(0,0){\epsfxsize=3.5in\epsffile{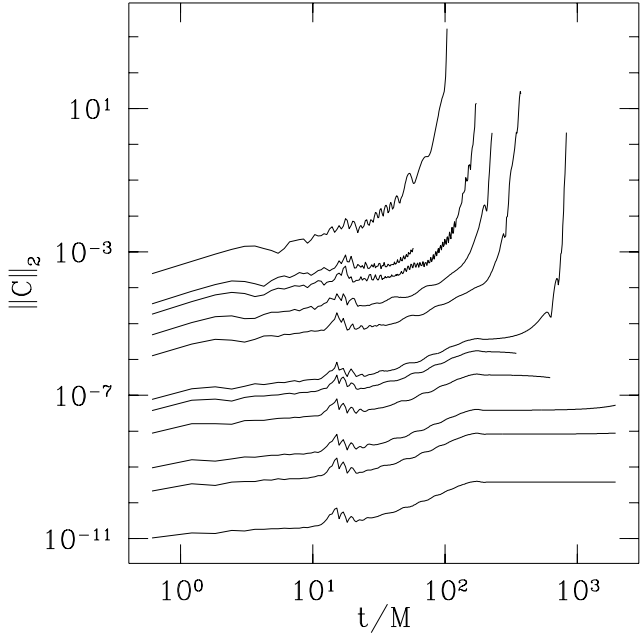}}
\end{picture}
\end{center}
\caption{Norm of the Hamiltonian constraint~(\ref{eq:hamcon}) as a
function of time for several spatial resolutions for evolutions of
harmonic initial data, run~\ref{WF04-14-2000B} of
Table~\ref{tableofruns}. The evolutions are identical to those in
Fig.~\ref{fig:WF04-17-2000DECHl2} except a backwards Euler time
stepping scheme is used.}
\label{fig:WF04-14-2000BECHl2}
\end{figure}

\begin{figure}
\begin{center}
\begin{picture}(240,240)
\put(0,0){\epsfxsize=3.5in\epsffile{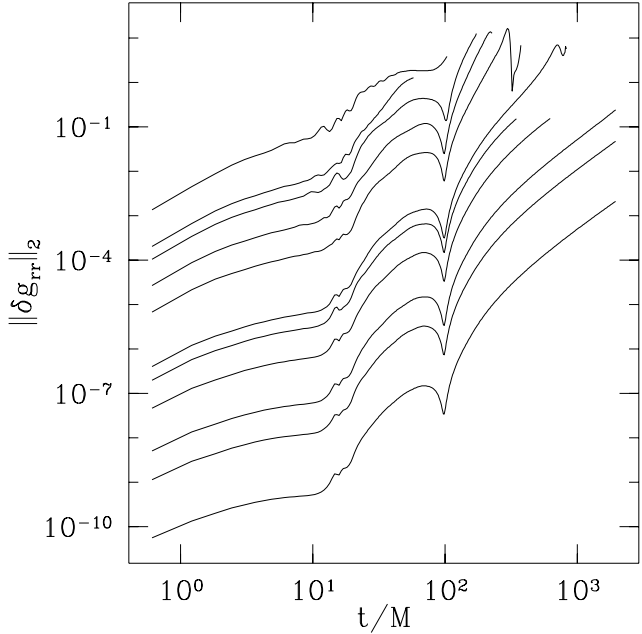}}
\end{picture}
\end{center}
\caption{Error in $g_{rr}$ versus time for the same evolutions shown
in Fig.~\ref{fig:WF04-14-2000BECHl2}. The growth is quadratic in $t$
at late times.}
\label{fig:WF04-14-2000Bgrrdiffl2}
\end{figure}

The $N_r=32$ case shown in Figs.~\ref{fig:WF04-17-2000DECHl2}
and~\ref{fig:WF04-17-2000Dgrrdiffl2} suffers from high-frequency noise
that grows exponentially in time. We have experimented with various
methods of damping this noise, including filtering the fundamental
variables after each time step and adding numerical dissipation terms
to the equations. However, we have obtained best results by changing
our fourth order Runge-Kutta time-stepping algorithm to an implicit
backwards Euler scheme, which is much more
dissipative. Figures~\ref{fig:WF04-14-2000BECHl2}
and~\ref{fig:WF04-14-2000Bgrrdiffl2} show the results of this
modification. The evolution now satisfies the constraints at late
times for sufficiently fine resolution, but still suffers from a
quadratically growing gauge mode that causes the coarser resolution
runs to crash. This gauge mode can be suppressed by applying active
gauge conditions, as shown in Section~\ref{sec:Ellipt-gauge-cond}
below.  Evolutions of fully harmonic initial
data~(\ref{full_harmonic}) produce results similar to those shown in
Figs.~\ref{fig:WF04-18-2000GECHl2}--\ref{fig:WF04-14-2000Bgrrdiffl2}.

We note that even with analytic gauge conditions and freezing outer
boundary conditions, evolutions of harmonic and fully harmonic initial
data such as those shown in Fig.~\ref{fig:WF04-18-2000GECHl2} become
stable when the outer boundary is moved sufficiently close to the
black hole (see runs~\ref{WF04-10-2000} and~\ref{O04-26-2000H}). A
similar dependence on the outer boundary location has also been
reported by others\cite{scheel_etal97b,Lehner2000}.  A possible
explanation for this is discussed briefly in\cite{scheel_etal97b}: For
a nonzero shift vector, any unstable zero-speed modes present in the
solution will propagate inward from the outer boundary with speed
$-\beta^r$, growing as they propagate. If the domain is sufficiently
small, these modes do not have time to grow appreciably before they
are swallowed by the horizon.  As discussed previously, we find that
constraint boundary conditions suppress exponentially growing modes,
and thus allow evolutions with a larger outer boundary radius
(runs~\ref{LKO05-10-2000}, \ref{LKO05-10-2000E}, \ref{LKO05-10-2000C}
and~\ref{LKO05-12-2000B}).


\subsubsection{Elliptic gauge conditions}
\label{sec:Ellipt-gauge-cond}

Although choosing a time-independent $\beta^r$ and $\tilde\alpha$ is
the simplest gauge condition to implement, for an evolving numerical
solution such a choice does not actively enforce any particular
coordinate condition.  In fact, it is remarkable that many of the
cases discussed in Section~\ref{sec:Analyt-gauge-cond} remain stable
when the coordinates experience small deviations from the exact
solution. For more than one black hole in three spatial dimensions,
one will almost certainly need general gauge conditions designed to
prevent large changes in the numerical solution of a stationary or
quasi-stationary spacetime.

\begin{figure}
\begin{center}
\begin{picture}(240,240)
\put(0,0){\epsfxsize=3.5in\epsffile{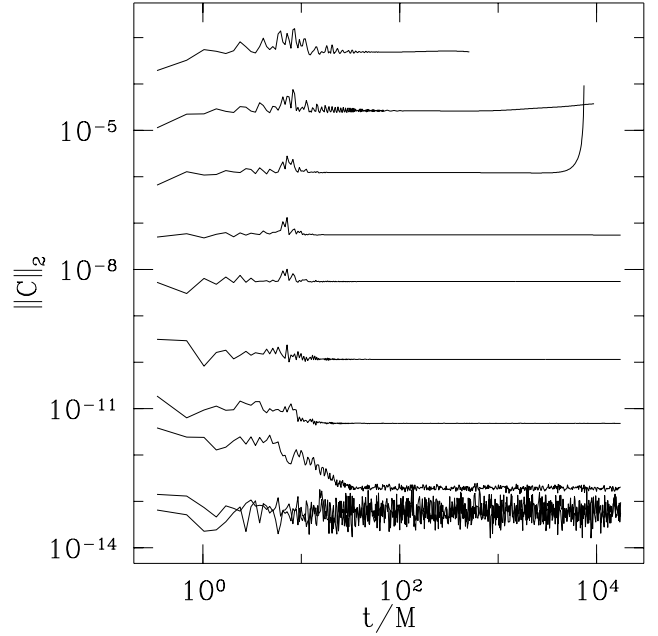}}
\end{picture}
\end{center}
\caption{Norm of the Hamiltonian constraint~(\ref{eq:hamcon}) as a
function of time for several spatial resolutions $N_r$ for evolutions
of Painlev\'{e}-Gullstrand initial data using elliptic gauge
conditions, run~\ref{LKWF04-20-2000A} of Table~\ref{tableofruns}.  The
number of spectral coefficients $N_r$ for each plot, starting at the
top, is $12$, $16$, $20$, $24$, $27$, $32$, $36$, $40$, $45$, and
$48$.}
\label{fig:WF04-20-2000AECHl2}
\end{figure}

\begin{figure}
\begin{center}
\begin{picture}(240,240)
\put(0,0){\epsfxsize=3.5in\epsffile{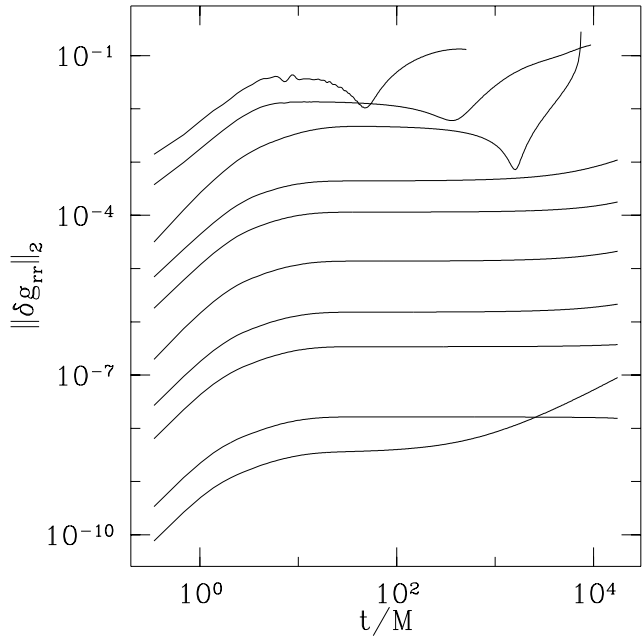}}
\end{picture}
\end{center}
\caption{Error in $g_{rr}$ versus time for the same evolutions
shown in Fig.~\ref{fig:WF04-20-2000AECHl2}. For the highest
resolution, the growth is only linear in $t$ at late times.}
\label{fig:WF04-20-2000Agrrdiffl2}
\end{figure}

Figure~\ref{fig:WF04-20-2000AECHl2} shows the norm of the Hamiltonian
constraint for an evolution of Painlev\'{e}-Gullstrand initial
data. The gauge variables $\beta^r$ and $\tilde\alpha$ are computed by
solving the minimal strain and stationary mean curvature
equations~(\ref{eq:MinStrain1D}) and~(\ref{eq:StatMeanCurv1D}) after
each time step. These elliptic equations require boundary
conditions. We impose~(\ref{eq:1DEC_freezehorizon})
and~(\ref{eq:cminuscondition}) at the current location of the horizon,
which we recompute after every time step.
For~(\ref{eq:cminuscondition}) we choose $c_{-}=-2$, which is the
value of $c_{-}$ at the horizon for the analytic
solution~(\ref{pain-gull}).  At the outer boundary, we set
$\tilde\alpha=1$ and we impose a Robin condition~(\ref{eq:Robin}) on
$\beta^r$ with $\beta^r_\infty=0$, $n=1/2$. As seen in the figure, the
evolution remains stable and convergent. To achieve stability, we find
it necessary to apply a simple $2/3$ cutoff filter to $\tilde\alpha$
and $\beta^r$ each time they are computed, and to impose constraint
boundary conditions on $U^0_r$ and $U^0_t$ (but not on
$U^-_T$).  Figure~\ref{fig:WF04-20-2000Agrrdiffl2} shows the
error in $g_{rr}$ for the same evolution.  For the highest resolution,
one can see a linearly-growing gauge mode.  Although modes that grow
linearly will eventually terminate a simulation, they pose no
difficulty for long-term evolutions because a much longer run time can
be achieved by a modest increase in resolution.

\begin{figure}
\begin{center}
\begin{picture}(240,240)
\put(0,0){\epsfxsize=3.5in\epsffile{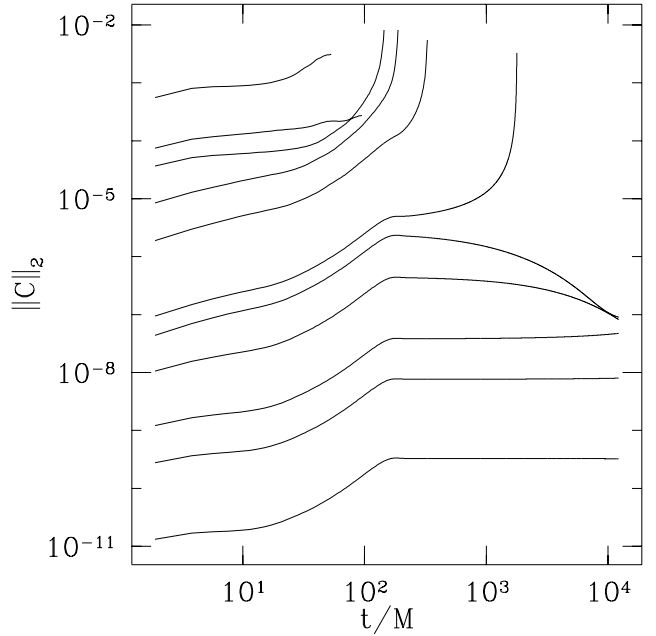}}
\end{picture}
\end{center}
\caption{Norm of the Hamiltonian constraint~(\ref{eq:hamcon}) as a
function of time for several spatial resolutions for evolutions of
harmonic initial data using elliptic gauge conditions,
run~\ref{WF04-18-2000E} of Table~\ref{tableofruns}. Resolutions are
the same as in Fig.~\ref{fig:WF04-18-2000GECHl2}.}
\label{fig:WF04-18-2000EECHl2}
\end{figure}

\begin{figure}
\begin{center}
\begin{picture}(240,240)
\put(0,0){\epsfxsize=3.5in\epsffile{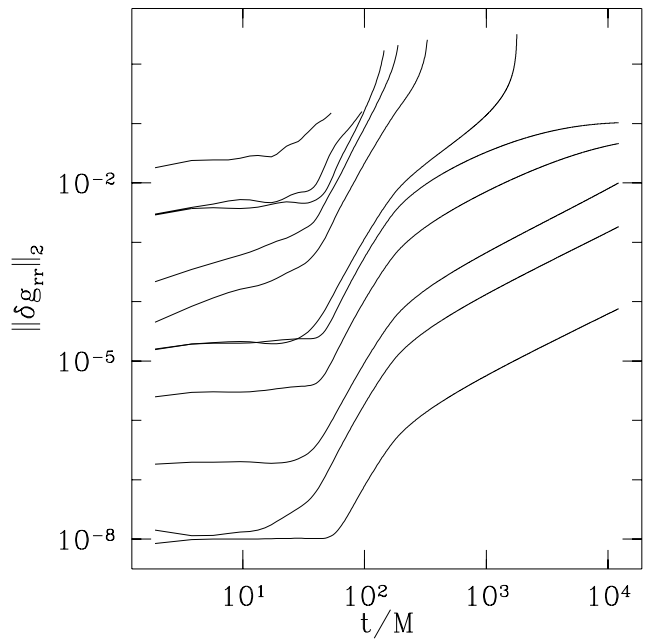}}
\end{picture}
\end{center}
\caption{Error in $g_{rr}$ versus time for the same evolutions 
shown in Fig.~\ref{fig:WF04-18-2000EECHl2}. The
growth is only linear in $t$ at late times.}
\label{fig:WF04-18-2000Egrrdiffl2}
\end{figure}

Similar results for the case of harmonic initial data are shown in
Figs.~\ref{fig:WF04-18-2000EECHl2}
and~\ref{fig:WF04-18-2000Egrrdiffl2}. The evolution is stable and
convergent, and the rapidly-growing gauge mode that terminated the
simulation in the case of time-independent $\beta^r$ and
$\tilde\alpha$ (Section~\ref{sec:Analyt-gauge-cond},
Fig.~\ref{fig:WF04-14-2000Bgrrdiffl2}) now grows only linearly with
time.  As in the case shown in
Fig.~\ref{fig:WF04-14-2000Bgrrdiffl2}, we use a backwards Euler
scheme for time evolution. For a fourth-order Runge-Kutta time
discretization, results are similar except the evolutions with
$N_r=25$, $30$, and $32$ are unstable.

\subsection{Black hole plus scalar wave}

In this subsection, we add dynamics to our spherically symmetric
spacetime by including a Klein-Gordon scalar field as a matter source,
as described in Section~\ref{sec:Einst-Klein-Gord}.  To set up initial
data, we first choose arbitrary initial values of $\Pi$ and $\Phi_r$.
We then solve the Hamiltonian and momentum constraints via the
standard York-Lichnerowicz conformal decomposition\cite{york79}, using
one of the time-independent representations of the Schwarzschild
geometry discussed in Section~\ref{sec:Time-indep-slic} as a
background solution.  Once we obtain the conformal factor and the
trace-free longitudinal part of the extrinsic curvature from the
constraints, we reconstruct the fundamental variables.

For the evolutions described here, the scalar field $\Pi$ is initially
a Gaussian centered at $r=20M$ with a width of $5M$ and an amplitude
of $0.02/M$, and $\Phi_r$ is zero everywhere. The outer boundary is
located at $r=120M$ and the inner boundary is at $1.75M$. Here $M$
is the mass of the background solution, which is different than the
actual mass of the spacetime.  We choose a Kerr-Schild background,
analytic gauge conditions, and a fourth-order Runge-Kutta time
stepping algorithm. At the outer boundary we apply freezing boundary
conditions~(\ref{eq:ConstantOuterBoundary}) to $U^-_\phi$ and $U^-_r$,
and constraint boundary conditions to $U^0_r$, $U^0_t$ and $U^-_T$.
There is no boundary condition imposed at the inner boundary.

To demonstrate our ability to handle multiple domains, for this
evolution we cover the entire domain with $8$ equal-sized abutting
subdomains, each using $45$ spectral coefficients. At each domain
boundary, the incoming characteristic quantities in each domain are
set equal to the corresponding outgoing quantities of the neighboring
domain.

\begin{figure}
\begin{center}
\begin{picture}(240,240)
\put(0,0){\epsfxsize=3.5in\epsffile{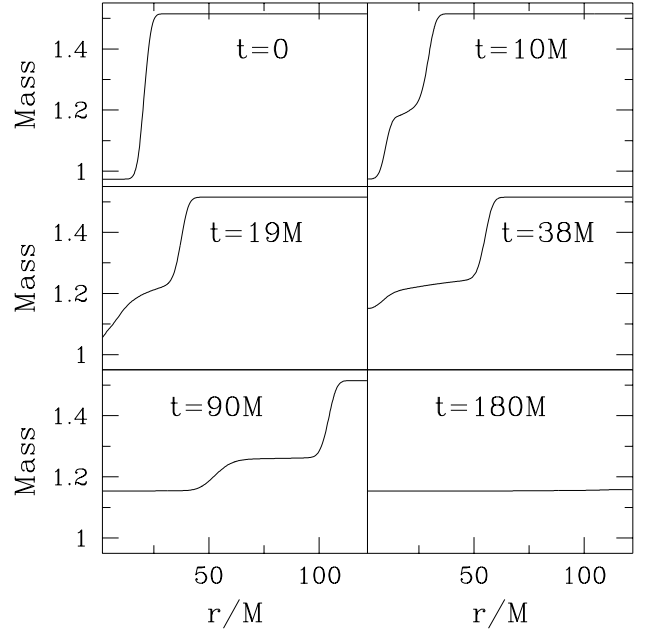}}
\end{picture}
\end{center}
\caption{Misner-Sharp mass, measured in units of $M$, as a function of
radius at selected times for an evolution of the Einstein-Klein-Gordon
system. Here $M$ is the mass of the background solution. See text for
details.}
\label{fig:ScalarMassEvolution}
\end{figure}

In Fig.~\ref{fig:ScalarMassEvolution} we plot the mass contained
within radius $r$ as a function of $r$ for selected times.  Initially
the mass of the black hole is $0.97M$ and the mass of the entire
spacetime is $1.52M$. The scalar field energy concentrated near
$r=20M$ accounts for $0.55M$. As the evolution proceeds, the initial
Gaussian scalar field pulse divides into incoming and outgoing pieces.
The outgoing piece propagates to infinity, while the incoming piece is
partially reflected off the Schwarzschild potential and partially
swallowed by the black hole.  At $t=90M$ the mass of the black hole
has reached its final value of $1.15M$, and the initial outgoing pulse
and the reflected pulse have not yet reached the outer boundary of the
domain. By $t=180M$ the remaining scalar radiation has left the
domain. 

\begin{figure}
\begin{center}
\begin{picture}(240,240)
\put(0,0){\epsfxsize=3.5in\epsffile{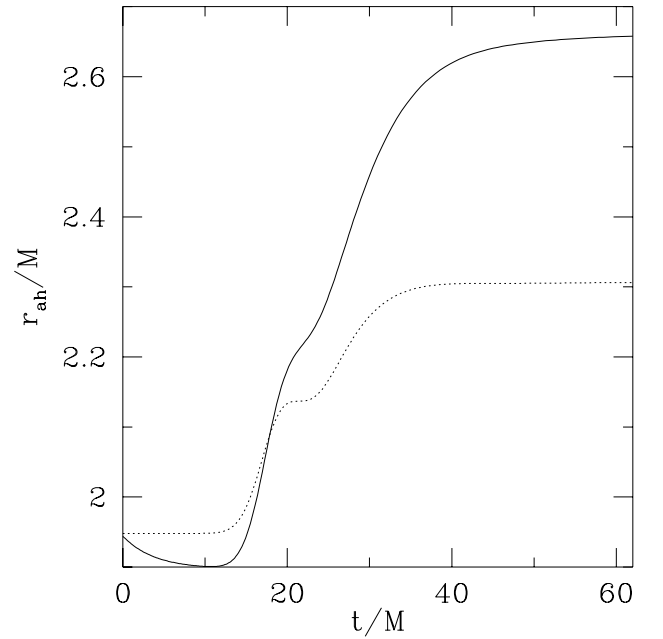}}
\end{picture}
\end{center}
\caption{Coordinate radius (solid line) and areal radius (dotted line)
of the apparent horizon as a function of time for the evolution shown
in Fig.~\ref{fig:ScalarMassEvolution}.}
\label{fig:ScalarWaveHorizon}
\end{figure}

Figure~\ref{fig:ScalarWaveHorizon} shows the coordinate radius and the
areal radius of the apparent horizon versus time. The area of the
horizon increases between $t=15M$ and $t=30M$ as the scalar field
pulse falls into the black hole, and the areal radius asymptotically
approaches the value $2.31M$, which is twice the mass of the final
black hole as expected.  Unlike the areal radius, the coordinate
radius decreases with time until $t=10M$, after which it increases and
eventually asymptotes to $r=2.66M$.  One can see from the figure that
on the initial slice the coordinate $r$ is nearly areal, as it would
be for the Kerr-Schild background solution without a scalar field. At
late times the deviation from an areal radial coordinate is large.

\begin{figure}
\begin{center}
\begin{picture}(240,240)
\put(0,0){\epsfxsize=3.5in\epsffile{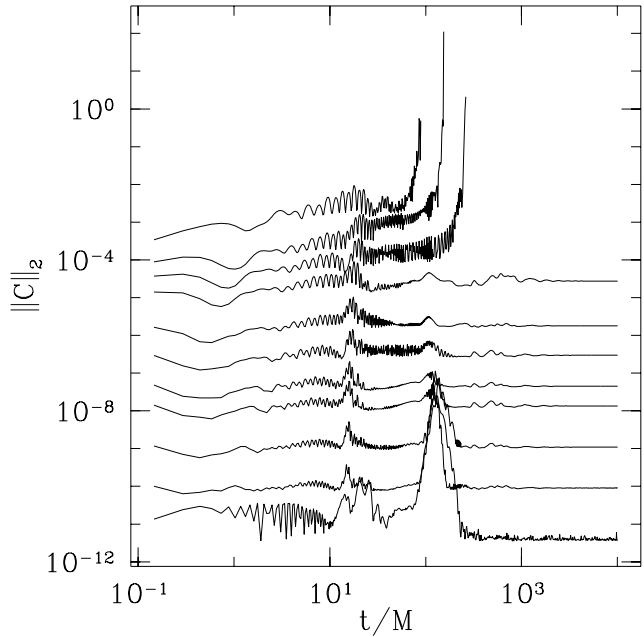}}
\end{picture}
\end{center}
\caption{Norm of the Hamiltonian constraint~(\ref{eq:hamcon}) as a
function of time for the evolution shown in
Fig.~\ref{fig:ScalarMassEvolution} and for coarser evolutions with
$12$, $16$, $18$, $20$, $24$, $27$, $30$, $32$, $36$, and $40$ 
spectral coefficients per domain.}
\label{fig:ScalarECHl2}
\end{figure}

\begin{figure}
\begin{center}
\begin{picture}(240,240)
\put(0,0){\epsfxsize=3.5in\epsffile{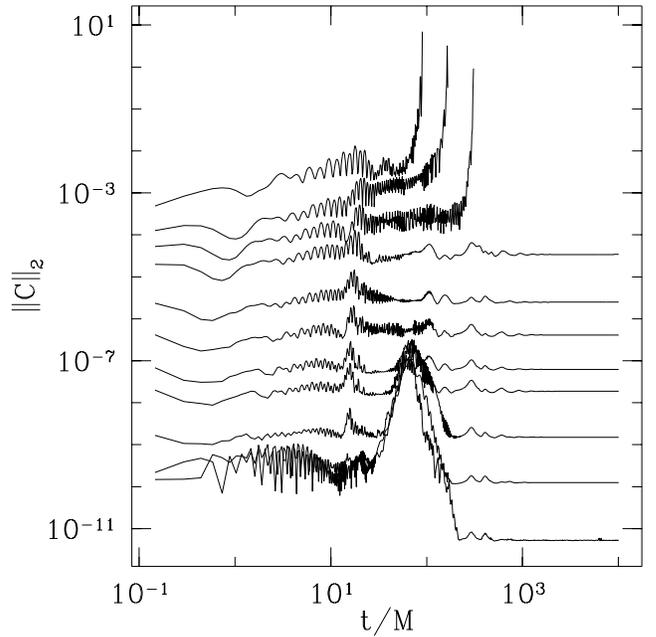}}
\end{picture}
\end{center}
\caption{Same as Fig.~\ref{fig:ScalarECHl2} except each
evolution has an outer boundary radius of $60M$ instead of $120M$.}
\label{fig:ScalarECHl2_60M}
\end{figure}

Figure~\ref{fig:ScalarECHl2} shows the norm of the Hamiltonian
constraint as a function of time for the same evolution, as well as
for other evolutions with different spatial resolutions but the same
$\Delta t$.  The plots exhibit exponential convergence, even with
nontrivial dynamics and multiple domains.  There is, however, a small
failure of convergence in the highest resolutions around $t=120M$.
This is because the boundary
condition~(\ref{eq:ConstantOuterBoundary}), which is applied to
$U^-_r$ and $U^-_\phi$ at the outer boundary $r_b$, is not strictly
correct while a wave is passing through the boundary, and the error
this introduces scales like ${r_b}^{-2}$.  We have verified this by
repeating the evolutions from Fig.~\ref{fig:ScalarECHl2} with the
inner boundary a factor of two closer, at $60M$. This is shown in
Fig.~\ref{fig:ScalarECHl2_60M}. The resolution per subdomain is the
same as in Fig.~\ref{fig:ScalarECHl2}, but we use $4$ equal-sized
subdomains instead of $8$.  The small nonconvergent feature in
Fig.~\ref{fig:ScalarECHl2_60M} is approximately a factor of four
larger than in Fig.~\ref{fig:ScalarECHl2}, and occurs at an earlier
time, $t=60M$, because the wave pulse reaches the outer boundary a
factor of two earlier.

\section{Discussion}
\label{sec:Discussion}

We have found that, at least for spherical symmetry, applying PSC
methods to hyperbolic formulations of general relativity can achieve
stable evolutions of black holes with horizon excision. Excision
itself is trivial as long as one uses a formulation in which all
characteristic speeds are causal.  Using realistic elliptic gauge
conditions, our evolutions are limited only by linearly growing gauge
modes that converge exponentially to zero with increasing resolution.
These modes create no difficulty for long-term simulations because a
small increase in resolution enables one to run much farther in time.
We note that even when errors grow exponentially in time (e.g.,
Fig.~\ref{fig:WF04-18-2000GECHl2}), the high accuracy provided by
PSC allows us to evolve to times of hundreds or sometimes thousands of
$M$.

A hyperbolic formulation of Einstein's equations provides a
straightforward way in which to formulate and implement boundary
conditions using the complete set of characteristic eigenfields
provided by hyperbolicity.  In principle, our method can be applied to
any hyperbolic formulation, but so far we have only used the EC
system.  We have not investigated whether PSC can be used with
non-hyperbolic formulations of Einstein's equations such as ADM, as
this would require a different treatment of the boundary conditions.
It is entirely possible, however, that one might find boundary
conditions that result in stable evolutions for such a formulation.

There has also been some concern about using hyperbolic
representations of general relativity with complicated gauge
conditions. This is because hyperbolic formulations of Einstein's
equations formally require the gauge quantities (shift and densitized
lapse in the case of EC) to be prescribed functions of space and time,
and not evolved quantities that couple to the fundamental variables.
However, as long as the gauge variables are held fixed during each
entire time step, as discussed by \cite{Yorkpc} and \cite{Bona1997},
we find no fundamental difficulty in applying elliptic gauge
conditions during our simulations.

By using the constraints as boundary conditions on the hyperbolic
evolution equations, we have found that one can improve evolutions of
the EC system.  We have found similar improvement for
finite-difference evolutions as well. Even in the general 3D case,
applying constraint boundary conditions on the metric variables is
straightforward. However, casting the Hamiltonian and momentum
constraints as boundary conditions may be more difficult, because
unlike in spherical symmetry, these constraints involve contractions
of derivatives of fundamental variables.

\begin{figure}
\begin{center}
\begin{picture}(240,120)(0,0)
\put(0,0){\epsfxsize=3.5in\epsffile{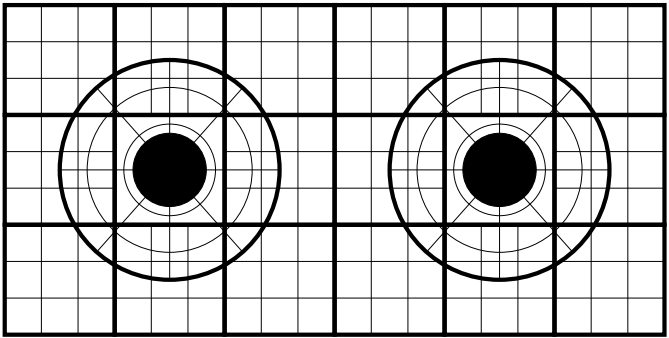}}
\end{picture}
\end{center}
\caption{Two-dimensional illustration of multiple computational
domains that might be used
to solve the binary black hole problem. Each hole is
surrounded by a single domain in the shape of a spherical
shell. Multiple cubical domains
overlap the spherical shells.
There are sixteen shown here in two dimensions, with
the cubes containing the black holes excised.}
\label{fig:multidomain}
\end{figure}

The numerical techniques discussed in this paper should be
generalizable to three spatial dimensions. For PSC evolutions of two
black holes with excised horizons it will be necessary to use multiple
computational domains (see Fig.~\ref{fig:multidomain}).  In
spherical symmetry, we have shown how multiple domains can be easily
implemented in a natural way by using characteristic fields to provide
inter-domain boundary conditions.  This method is directly applicable
to abutting domains in 3D, and the extension from abutting domains to
overlapping domains is straightforward \cite{Kopriva1989}.  Work on 3D
black hole evolutions using PSC is in progress.

\begin{acknowledgements}

This work was supported in part by NSF grants PHY-9800737 and
PHY-9900672 and NASA Grant NAG5-7264 to Cornell University, and NSF
grants PHY-9802571 and PHY-9988581 to Wake Forest
University. Computations were performed on the National Computational
Science Alliance SGI Origin2000, and on the Wake Forest University
Department of Physics IBM SP2 with support from an IBM SUR grant.

\end{acknowledgements}



\end{document}